\documentclass[12pt]{article}
\pdfoutput=1
\usepackage[latin1]{inputenc}
\usepackage[title]{appendix}
\usepackage{amsfonts}
\usepackage{amssymb}
\usepackage{mathtools}
\usepackage{graphicx}
\usepackage{wrapfig}
\usepackage{caption}
\usepackage{color}
\usepackage{url}
\usepackage[colorlinks,linkcolor=blue]{hyperref}

\makeatletter

\begin{document}
\unitlength = 1mm

\title{\bf 3D Network Model for Strong Topological Insulator Transitions}

\author{ Jun Ho Son$^1$ and S. Raghu$^{1,2}$\\
[7mm] \\
{\normalsize \it $^{1}$Stanford Institute for Theoretical Physics, Stanford 
University, Stanford, CA 94305, USA} \\
{\normalsize \it $^{2}$ Stanford Institute for Materials and Energy  Sciences, SLAC National} \\ 
{\normalsize \it Accelerator Laboratory, Menlo Park, CA 94025, USA}
}

\maketitle

\begin{abstract}
We construct a three-dimensional (3D), time-reversal symmetric  generalization of the Chalker-Coddington network model for the integer quantum Hall transition. The novel feature of our network model is that in addition to a weak topological insulator phase already accommodated by the network model framework in the pre-existing literature, it hosts strong topological insulator phases as well. We unambiguously demonstrate that strong topological insulator phases emerge as intermediate phases between a trivial insulator phase and a weak topological phase. Additionally, we found a  non-local transformation that relates a trivial insulator phase and a weak topological phase in our network model. Remarkably, strong topological phases are mapped to themselves under this transformation. We show that upon adding sufficiently strong disorder the strong topological insulator phases undergo phase transitions into a metallic phase. We numerically determine the critical exponent of the insulator-metal transition. Our network model explicitly shows how a semi-classical percolation picture of topological phase transitions 
in 2D can be generalized to 3D and opens up a new venue for studying 3D topological phase transitions.
\end{abstract}

\newpage

\tableofcontents

\section{Introduction}

 The Chalker-Coddington network model \cite{Chalker1988} has been a popular choice of models for studying non-interacting integer quantum Hall phase transitions with quenched disorder in two spatial dimensions. This model provides an intriguing semiclassical picture for the integer quantum Hall transitions: Systems with random but smooth chemical potentials are viewed as systems with ``mesoscopic" droplets of local integer quantum hall phases in background of a topologically trivial insulator, mesoscopic droplets interpreted as local valleys of chemical potential. The Chalker-Coddington model strips away all degrees of freedom except those from edge modes between the trivial background and the droplets. Tunneling between edge modes in neighboring droplets induces topological phase transitions. The critical point occurs when the edge modes ``percolate" throughout the whole network.  
 
 Originally devised for the integer quantum Hall transition (class A in the Altland-Zirnbauer classification \cite{Altland1997,Ryu2010}), the network model has been generalized to all other symmetry classes in which there are non-trivial topological phases \cite{Cho1997,Gruzberg1999,Beamond2002, Obuse2007,Fulga2012}; the generalized network model in each symmetry class describes critical phenomena between two topologically inequivalent gapped phases. Also, the network model was generalized into phase transitions involving a trivial insulator and a 3D \textit{weak} topological insulator \cite{Chalker1995}. However, to our knowledge, a construction of a 3D network model whose phase diagram includes a \textit{strong time-reversal invariant topological insulator} does not exist.

 Our goal here is to construct such a 3D time-reversal invariant network model. In Sec.~\ref{sec:model}, we will motivate and construct a 3D network model that can describe topological phase transitions involving 3D strong topological insulators and point out some interesting dualities in the model. This section will make manifest how the semiclassical percolation picture of the 2D Chalker-Coddington model can be generalized to 3D. In Sec.~\ref{sec:phase}, we present a detailed study of phase diagrams, both with and without quenched randomness. While the phase diagram for the clean system is largely a quantitative confirmation of what we anticipated in Sec.~\ref{sec:model}, the phase diagram for the dirty system has more intriguing features, most notably exotic phase transitions from strong topological insulators to metallic phases induced by \textit{increasing disorder} rather than decreasing disorder as expected from the conventional localization-delocalization transition.

\section{Model}
\label{sec:model}
\subsection{Review of the 2D Chalker-Coddington Network model}

 In this subsection, we briefly review the Chalker-Coddington model \cite{Chalker1988} and its time-reversal symmetric generalization \cite{Obuse2007} to motivate our 3D construction and to set up some key notation.
 
 \subsubsection{Chalker-Coddington Model}
 In the Chalker-Coddington model, degrees of freedom are represented as complex number variables $\psi_{(x,y)}$ associated with wavefunction amplitudes of fermionic states at site $(x,y)$. For odd $x$,  $\psi_{(x,y)}$ is associated with fermions moving in $+y$ direction; $\psi_{(x,y)}$ with even $x$ is associated with fermions moving in $-y$ directions. Hence, fermion modes have chirality structure built in the setup. Additionally, there are local unitary scattering matrices that endow local relations among neighboring $\psi_{(x,y)}$'s. 
 
\begin{figure}
\includegraphics[width=0.9\linewidth]{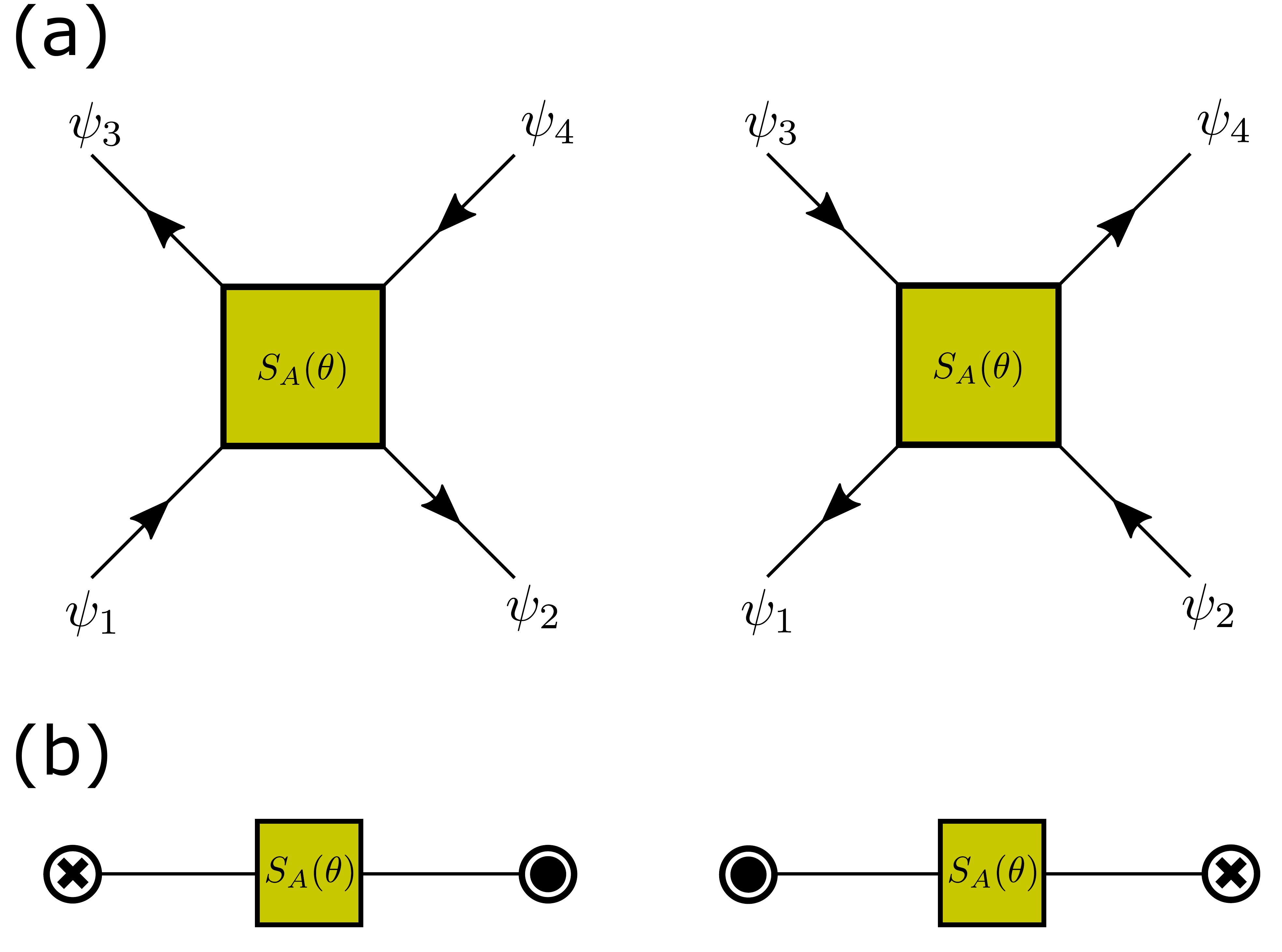}
\caption{The illustration of $S_{A}(\theta)$ in Eq.~\eqref{eq:2DclassAsmat} (a) in the bird's eye view and (b) in the cross-sectional view.}
\label{fig:Asmat}
\end{figure}

 A $2 \times 2$ scattering matrix that act as a basic building block for the full network, in the setup illustrated in either left or right of Fig.~\ref{fig:Asmat}(a), is the following:
\begin{equation}
\label{eq:2DclassAsmat}
S_{A}(\theta) \begin{pmatrix}
\psi_{1} \\
\psi_{4}
\end{pmatrix} = \begin{pmatrix}
\psi_{2} \\
\psi_{3}
\end{pmatrix}, \quad S_{A}(\theta) = \begin{pmatrix}
i\cos{\theta} & \sin{\theta}\\
\sin{\theta} & i\cos{\theta}
\end{pmatrix}
\end{equation}
 The key idea is that $\psi_{1}$ and $\psi_{4}$ ($\psi_{2}$ and $\psi_{3}$) are associated with incoming modes (outgoing modes) on the left side of Fig.~\ref{fig:Asmat}(a); the role of incoming and outgoing modes are reversed on the right side of Fig.~\ref{fig:Asmat}(a). In either cases, $S_{A}(\theta)$ unitarily relates two incoming modes and two outgoing modes. 
 
 For the sake of later convenience when we introduce our 3D model, we also adopt cross-sectional graphical illustrations of $S_{A}(\theta)$ in Fig.~\ref{fig:Asmat}(b). The key observation behind this graphical notation is that the pairs $(\psi_{1},\psi_{3})$ and $(\psi_{2},\psi_{4})$ move in opposite vertical directions from each other. Hence, we utilize $\otimes$ and $\odot$ symbols familiar from introductory electromagnetism courses to represent these pairs in the cross-sectional view.
 
\begin{figure}
\includegraphics[width=1.0\linewidth]{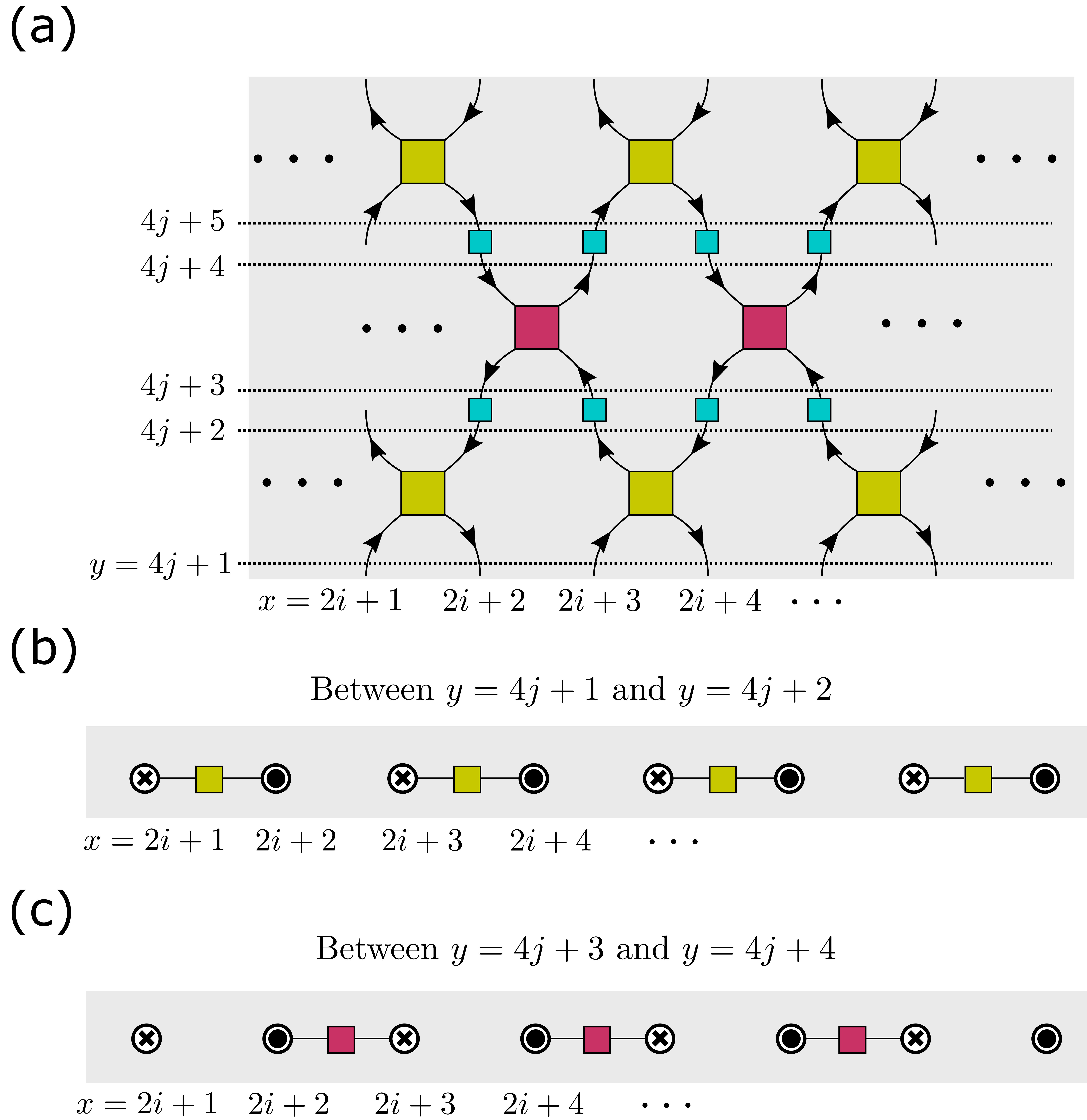}
\caption{The illustration of the Chalker-Coddington model, (a) from the bird's eye view (b,c) views from the cross sections.}
\label{fig:CCarray}
\end{figure}

 The Chalker-Coddington model has one tuning parameter, which we call $\theta_{CC}$. There are three types of local relations in the construction of the network model: scattering matrices $S_{A}(\theta_{CC})$ that relate fermion amplitudes $\psi_{(x,y)}$ with $x=2i+1, 2i+2$ and $y =4j+1, 4j+2$ (gold squares in Fig.~\ref{fig:CCarray}), scattering matrices $S_{A} \left( \frac{\pi}{2}  - \theta_{CC} \right) $  that relate fermion amplitudes $\psi_{(x,y)}$ with $x=2i+2, 2i+3$ and $y =4j+3, 4j+4$ (red squares in Fig.~\ref{fig:CCarray}), and blue squares in Fig.~\ref{fig:CCarray} between $y=4j+2$ and $y=4j+3$ and between $y=4j+4$ and $y=4j+5$. Blue squares between $(x,y)$ and $(x,y+1)$ induce the following relation:
\begin{equation}
\psi_{(x,y)} = e^{i\phi_{(x,y)}} \psi_{(x,y+1)}
\end{equation}
Where $ e^{i\phi_{(x,y)}}$ is a random U$(1)$ phase. Hence, upon crossing the blue squares fermion modes acquire random U$(1)$ phases, and these random phases encode the effect of disorder. 
Fig.~\ref{fig:CCarray}(a) shows how these components are arranged in the network model. Alternatively, one can use the cross-sectional view for the more compact illustration of how the scattering matrices are placed, as in Fig.~\ref{fig:CCarray}(b) and (c).
 
From a numerical standpoint, thanks to how scattering matrices are placed in the model, one can naturally construct $\mathbf{S}_{y}$, a scattering matrix that relates fermion amplitudes at $y$ and $y+1$ from local scattering matrices illustrated in Fig.~\ref{fig:CCarray}, for each $y$. Then, one may use $\mathbf{S}_{y}$'s to build a global scattering matrix $\mathbf{S}$ that relates $\psi$'s at the start of the network and $\psi$'s at the end of the network in an effective manner. From this global scattering matrix, one can extract information such as  Landauer conductance and topological invariants \cite{Fulga2012b}, allowing one to study phase diagrams and critical properties \cite{Abrahams1979} in detail. Alternatively, one can convert scattering matrices $\mathbf{S}_{y}$'s to transfer matrices $\mathbf{T}_{y}$'s and use transfer matrices to compute Lyapunov exponents to study localization properties \cite{MacKinnon1981,MacKinnon1983}. In Appendix A, we discuss how to construct transfer matrices from scattering matrices and how the global scattering matrix can be efficiently evaluated numerically.
 
 While the Chalker-Coddington model is amenable to numerical methods, one can obtain a great deal of information, without explicit computation, by considering certain limits and invoking a duality that is inherent to the model. Before discussing phase diagrams, it is useful to have some intuition on $S_{A}(\theta)$ in Eq.~\eqref{eq:2DclassAsmat} from the cross-sectional view in Fig.~\ref{fig:Asmat}(b). If $\theta=\frac{\pi}{2}$, in the setup represented by Fig.~\ref{fig:Asmat}(b), the left modes and right modes represented as $\otimes$ and $\odot$ symbols are completely decoupled from each other, and $S_{A}(\theta)$ is in the perfect transmission limit. Meanwhile, when $\theta =0$, the left modes coming in always reflect to the right modes, and vice versa. Hence, this limit maximizes backscattering between the left modes and the right modes. Essentially, in Fig.~\ref{fig:Asmat}(b), tuning $\theta$ can be understood as tuning backscattering between fermion modes on the left and fermion modes on the right. 
 
 Having this intuition at hand, it is useful to consider $\theta_{CC}=0$ and $\theta_{CC} = \frac{\pi}{2}$ limit. At $\theta_{CC}=\frac{\pi}{2}$, the gold squares in Fig.~\ref{fig:CCarray}(b) are in the perfect transmission limit, but two modes coupled by a red square maximally backscatters with each other. This has an interesting consequence: When open boundary condition is imposed along $x$-direction, the chiral modes at $x=1$ and $x=2L_{x}$ ($L_{x}$ represents width of the model, so $x=2L_{x}$ is the largest $x$-index) remain completely decoupled from bulk degrees of freedom. Existence of chiral edge modes indicate that this phase should be identified as a $\nu=1$ quantum Hall phase. Meanwhile, at the limit $\theta_{CC}=0$, there is maximal backscattering induced between modes within a unit cell ($x=2i+1$ and $x=2i+2$), and such backscattering tunes the system into a trivial insulator.
 
 The following duality puts additional constraints on the phase diagram: Now consider imposing periodic boundary condition along $x$-direction, effectively making the system into a cylinder with circumference $2L_{x}$. Then, the substitution $\psi_{(x,y)} \rightarrow \psi_{(x+1 \mod 2L_{x} ,y)}$ maps the network with $\theta_{CC} = \theta_{0}$ to the network with $\theta_{CC} = \frac{\pi}{2} - \theta_{0}$. This duality relates the $\nu=1$ quantum Hall phase near $\theta_{CC} = \frac{\pi}{2}$ to the trivial insulator phase near $\theta_{CC} = 0$. Additionally, if the only two possible extended phases in the phase diagram are a $\nu=1$ quantum Hall state and a trivial insulator (This is a natural expectation given that the Chalker-Coddington model is essentially a 2D non-interacting fermion system with disorder), the critical point between two phases is fixed to be at the self-dual point $\theta_{CC} = \frac{\pi}{4}$. Hence, the network model has the self-duality at the quantum Hall transition by construction. 

\subsubsection{The Time-Reversal Symmetric Generalization}
 
 To construct the time-reversal symmetric generalization of the Chalker-Coddington model in \cite{Obuse2007}, we will add time-reversed partners to the degrees of freedom in the original model for the integer quantum Hall transition. Now, $\psi_{(x,y),s=\uparrow,\downarrow}$ carry spin labels in addition to spatial indices as those in the original Chalker-Coddington network model. $\psi_{(x,y),\uparrow}$ and $\psi_{(x,y),\downarrow}$ form a Kramer's pair. $\psi_{(x,y),\uparrow}$ may be understood as inherited from the original Chalker-Coddington model; especially, it has the same chirality structure as $\psi_{(x,y)}$ of the original model. In contrast, $\psi_{(x,y),\downarrow}$'s are associated with fermions flowing in the opposite direction from those represented by the $s=\uparrow$ counterpart. Under time-reversal symmetry, $\psi$'s transform as:
 
\begin{equation}
\mathcal{T}: \, \psi_{\uparrow,(x,y)} \rightarrow \psi_{\downarrow,(x,y)}^{*}, \quad \psi_{\downarrow,(x,y)} \rightarrow -\psi_{\uparrow,(x,y)}^{*}
\end{equation} 
 
\begin{figure}
\includegraphics[width=1.0\linewidth]{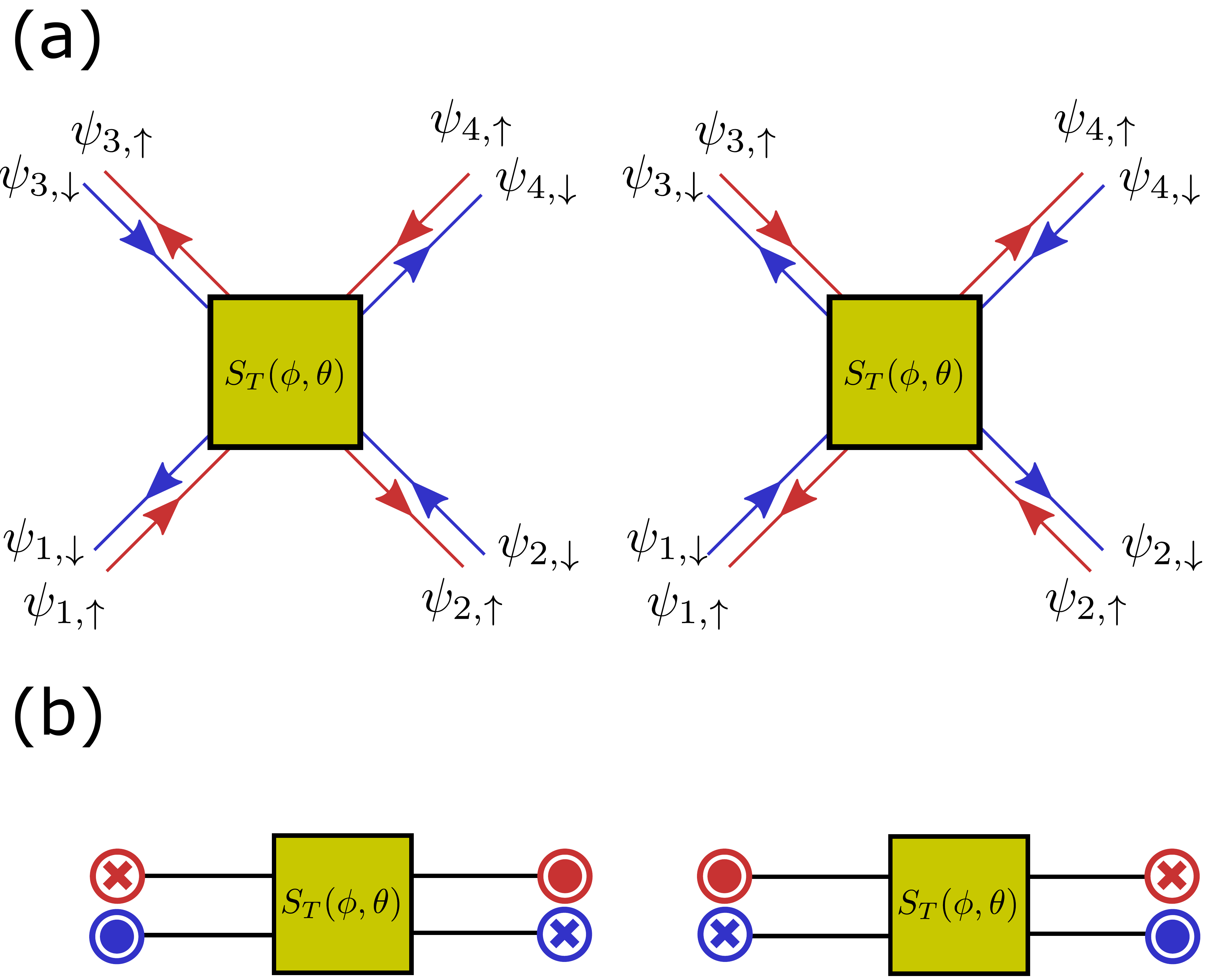}
\caption{(a) The setup and the labels used in the definition of Eq.~\eqref{eq:2DclassAIIsmat} (b) Cross-sectional view of (a). In both figures and in future figures as well, we use red lines/symbols for spin-up components and blue lines/symbols for spin-down components.}
\label{fig:AIIsmat}
\end{figure}

 Now, a basic building block of the time-reversal symmetric network, i.e., taking a role of $S_{A}(\theta)$ in Eq.~\eqref{eq:2DclassAsmat} in the Chalker-Coddington model, is a $4 \times 4$ scattering matrix $S_{T}(\phi, \theta)$:
\begin{equation}
 \label{eq:2DclassAIIsmat}
\begin{split}
S_{T}(\phi,\theta) \begin{pmatrix}
\psi_{1,\uparrow} \\
\psi_{2,\downarrow} \\
\psi_{3,\downarrow} \\
\psi_{4,\uparrow} \\
\end{pmatrix} = \begin{pmatrix}
\psi_{1,\downarrow} \\
\psi_{2,\uparrow} \\
\psi_{3,\uparrow} \\
\psi_{4,\downarrow} \\
\end{pmatrix}, \quad S_{T}(\phi,\theta) = \begin{pmatrix}
0 & i\cos{\phi}\cos{\theta} & \cos{\phi}\sin{\theta} & \sin{\phi} \\
i\cos{\phi}\cos{\theta} & 0 & \sin{\phi} & \cos{\phi}\sin{\theta} \\
\cos{\phi}\sin{\theta} & -\sin{\phi} & 0 &  i\cos{\phi}\cos{\theta}  \\
-\sin{\phi} & \cos{\phi}\sin{\theta} & i\cos{\phi}\cos{\theta} & 0  
\end{pmatrix} 
\end{split}
\end{equation} 
 See Fig.~\ref{fig:AIIsmat}(a) for the labels. We also adopt the cross-sectional illustration as in Fig.~\ref{fig:AIIsmat}(b) as well.

 The time-reversal symmetry requires that $S_{T}(\phi , \theta)$ should satisfy:
\begin{equation}
\label{eq:tconstraint}
-U_{\mathcal{T}} S_{T}^{*}(\phi , \theta) U_{\mathcal{T}} =  S_{T}^{\dagger}(\phi , \theta) , \quad U_{\mathcal{T}} = \begin{pmatrix}
1 & 0 & 0 & 0 \\
0 & -1 & 0 & 0 \\
0 & 0 & -1 & 0 \\
0 & 0 & 0 & 1
\end{pmatrix}
\end{equation}
One can explicitly check that the scattering matrix in Eq.\eqref{eq:2DclassAIIsmat} satisfies the above condition. 

 In the time-reversal symmetric network model, there are two tunable parameters: $\phi_{QSH}$ and $\theta_{QSH}$. The following modifications from the setup for the Chalker-Coddington model illustrated in Fig.~\ref{fig:CCarray} will yield the time-reversal symmetric network model: First, to account for doubled degrees of freedom, one should add the time-reversed partner modes moving in opposite directions to the ones already in the setup. Gold squares now represent $4 \times 4$ scattering matrices $S_{T}(\phi_{QSH}, \theta_{QSH})$, while red squares are now taken to be $S_{T}(\phi_{QSH}, \frac{\pi}{2}-\theta_{QSH})$. Finally, the blue square at $x$ and between $y$ and $y+1$ relates $\psi$'s as the following:
\begin{equation}
\label{eq:AIIrandom}
\begin{pmatrix}
e^{i\phi_{(x,y)}} & 0 \\
0 & e^{-i\phi_{(x,y)}}  \\
\end{pmatrix} \begin{pmatrix}
\psi_{(x,y),\uparrow} \\
\psi_{(x,y),\downarrow} \\
\end{pmatrix} = \begin{pmatrix}
\psi_{(x,y+1),\uparrow} \\
\psi_{(x,y+1),\downarrow} \\
\end{pmatrix} 
\end{equation}
Where $e^{i\phi_{(x,y)}}$ is a random U$(1)$ phase once again that encode the effect of disorder. Note that U$(1)$ phase that $s=\uparrow$ and $s=\downarrow$ modes acquire upon crossing a blue square is chosen to be exactly complex conjugate to each other so that time-reversal symmetry is satisfied.
 
 When $\phi_{QSH} = 0$, the spin-up sector and the spin-down sector are decoupled from one another. The model in this limit is simply two copies of the plain-vanilla network model stacked together and represents a physical system with $S_{z}$ conservation. Especially, $(\phi_{QSH}, \theta_{QSH}) = (0,0)$ corresponds to a trivial insulator upon following the logic presented in reviewing the phases of the original Chalker-Coddington model; in the case $(\phi_{QSH}, \theta_{QSH}) = (0,\frac{\pi}{2})$, the system is equivalent to $\nu=1$ and its time-reversed partner $\nu=-1$ phase stacked together, identified as a \textit{quantum spin Hall insulator}. $\phi_{QSH} \neq 0$ introduces coupling between modes with different spins. This coupling explicitly breaks $S_{z}$ conservation symmetry and allows one to access physics unique to the symplectic symmetry class, such as symplectic critical metal that exists as an \textit{extended phase} between a quantum spin Hall phase and a trivial phase while the system remains to be a quantum spin Hall insualtor (a trivial insulator) near $\theta_{QSH} = \frac{\pi}{2}$ ($\theta_{QSH} = 0$).
 
\subsection{Overview}

\begin{figure}
\includegraphics[width=\linewidth]{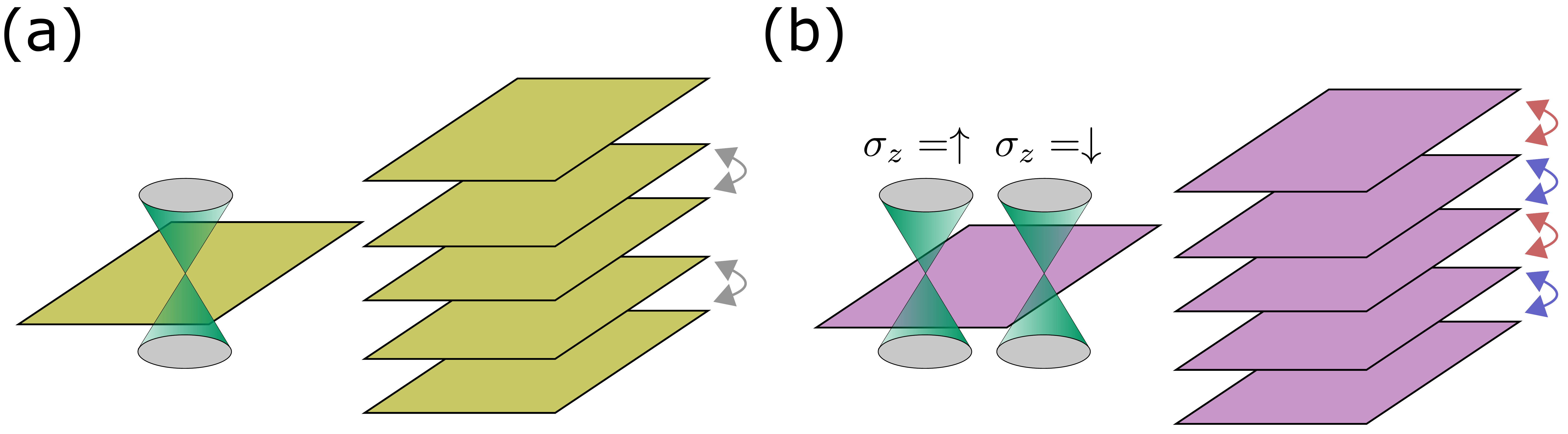}
\caption{(a) The naive generalization of the 2D approach to access a strong topological insulator phase from the 3D network model. Here, one stacks the hypothetical 2D layers that encode physics of a single Dirac cone and turns on interlayer backscattering between two layers in alternating pattern. (b) The approach explored in this paper. Each layer now has two Dirac cones, one for $\sigma_{z}=+1$ and one for $\sigma_{z}=-1$; half of the degrees of freedom, marked in red in the figure, backscatter between layer $2i+1$ and $2i+2$, while the other half, colored in blue, backscatter between layer $2i$ and $2i+1$, roughly speaking. } 
\label{fig:overview}
\end{figure}

 We will start discussion of our 3D network model by giving intuition behind our construction. In the 2D network model construction reviewed in the previous paragraph, one could build the network model by stacking spin-filtered helical fermions in which fermions move in opposite directions if they have opposite spins and induce backscattering appropriately. The most naive way to generalize this picture to describe the 3D strong topological insulator would be the following: Assume that there is a 2D network that describes a \textit{single, time-reversal symmetric Dirac cone}. Now, one can imagine introducing strong backscattering between the layer $2i$ and $2i+1$ to gap out this Dirac cone. Due to how we introduce backscattering, upon imposing open boundary conditions along the direction toward which we stack the layers, a single Dirac cone will be exposed and will serve as a surface state of the 3D strong topological insulator. See Fig.~\ref{fig:overview}(a) for the illustration.
 
 One may question existence of such 2D networks because of the fermion doubling problem. However, we note that the networks described by scattering matrices may avoid usual fermion doubling problems. One way to see this is to take the scattering matrices as discrete-time evolution operators instead of local relations between static fermion amplitudes. This alternative viewpoint allows one to interpret the network as a Floquet system which is known to evade some doubling problems \cite{Sun2018,Higashikawa2019}.  We also observe that in the course of reviewing the 2D network model, we clearly assumed that fermions of interest are chiral, which cannot be captured in a local 1D lattice Hamiltonian.
 
 The prescription given in the previous paragraph is certainly tantalizing, but we are not aware of any time-reversal symmetric networks whose scattering matrices describe a single Dirac cone; at present, we have not succeeded in constructing such a network. Instead, to access 3D strong topological insulators from the network model viewpoint, we will take a more roundabout route: We will imagine stacking 2D time-reversal symmetric network models that describes the physics of $\textit{two}$ Dirac cones, i.e., the 2D network model for the quantum spin Hall transition,  and figure out how to ``gap out" half of the degrees of the freedom.   
 
 To be more specific, let us start with stacks of the 2D networks described by scattering matrices given in Eq.~\eqref{eq:2DclassAIIsmat}, explicitly tuning $\theta = \frac{\pi}{4}$ and $\phi = 0$ to keep things maximally simple. In this setup two fermions with different $\sigma_{z}$ eigenvalues are independent, and each component individually describe the physics of a single Dirac cone. Now, to expose only ``half" of the layer when the open boundary condition is imposed, we will imagine introducing interlayer scattering in the following way: Half of the fermions strongly backscatter between layer $2i+1$ and $2i+2$, and the remaining half strongly backscatter between layer $2i$ and $2i+1$. Then, when one imposes open boundary condition, half of the fermions on the top and the bottom layer do not strongly backscatter to neighboring layers, while all other modes are gapped out by strong backscattering. Hence, roughly speaking only ``half" of the layer will constitute gapless edge states, yielding a strong topological insulator. See Fig.~\ref{fig:overview}(b) for the illustration. We note that a similar idea has been employed in the context of the wire construction to study 3D fractional topological insulators  \cite{Sagi2015}.
 
 There is one caveat in the above intuition. While the 2D quantum spin Hall phase in some limit reduces to a $\nu=1$ topological phase for $s=\uparrow$ stacked with its time-reversed partner forming a $\nu=-1$ state for $s=\downarrow$, such a limit does not exist for 3D strong topological insulators. Especially, to access physics of any 3D strong topological insulators, $\sigma_{z} = \uparrow$ component and $\sigma_{z}=\downarrow $ components should be mixed by inter-layer coupling. Hence, it is difficult to make rigorous claims about how surface states emerge as we spelled out somewhat hand-wavingly in the previous paragraph. Nevertheless the intuition turns out to be largely correct, and we do obtain a strong topological insulator in the 3D network model built from this insight.

\subsection{3D network model}

 Armed with the intuition, now we will explicitly construct our 3D network model whose phase diagram incorporates a 3D strong topological insulator phase. As before, we have complex-numbered variable $\psi_{(x,y,z), s=\uparrow,\downarrow,}$ at each site. Following the spirit of the 2D model, $\psi_{(x,y,z),\uparrow}$ with odd $x$ and $\psi_{(x,y,z),\downarrow}$ with even $x$ are associated with fermions moving in $+z$-direction, while $\psi_{(x,y,z),\downarrow}$ with odd $x$ and  $\psi_{(x,y,z),\uparrow}$ with even $x$ represent fermions flowing in $-z$-direction. Since the fully 3D illustration of the network can be confusing and difficult to visually understand, we will mostly use the view from the cross-section normal to $z$-direction, following the graphical notations similar to the ones showcased previously in Fig.~\ref{fig:Asmat}(b), Fig.~\ref{fig:CCarray}(b), and Fig.~\ref{fig:AIIsmat}(b).

 There are two scattering matrices that we will use as basic building blocks. The first is $S_{T}(\theta) = S_{T}(\phi=0, \theta)$ defined in Eq.~\eqref{eq:2DclassAIIsmat}. We use these scattering matrices to encode a pair of two-dimensional Dirac cones living on each layer, hence $\theta$ will be tuned to $\frac{\pi}{4}$. We note that  $\phi$ can be tuned to be zero because ``spin-orbit coupling" is present in our construction in a different type of scattering matrices.
 
\begin{figure}
\includegraphics[width=\linewidth]{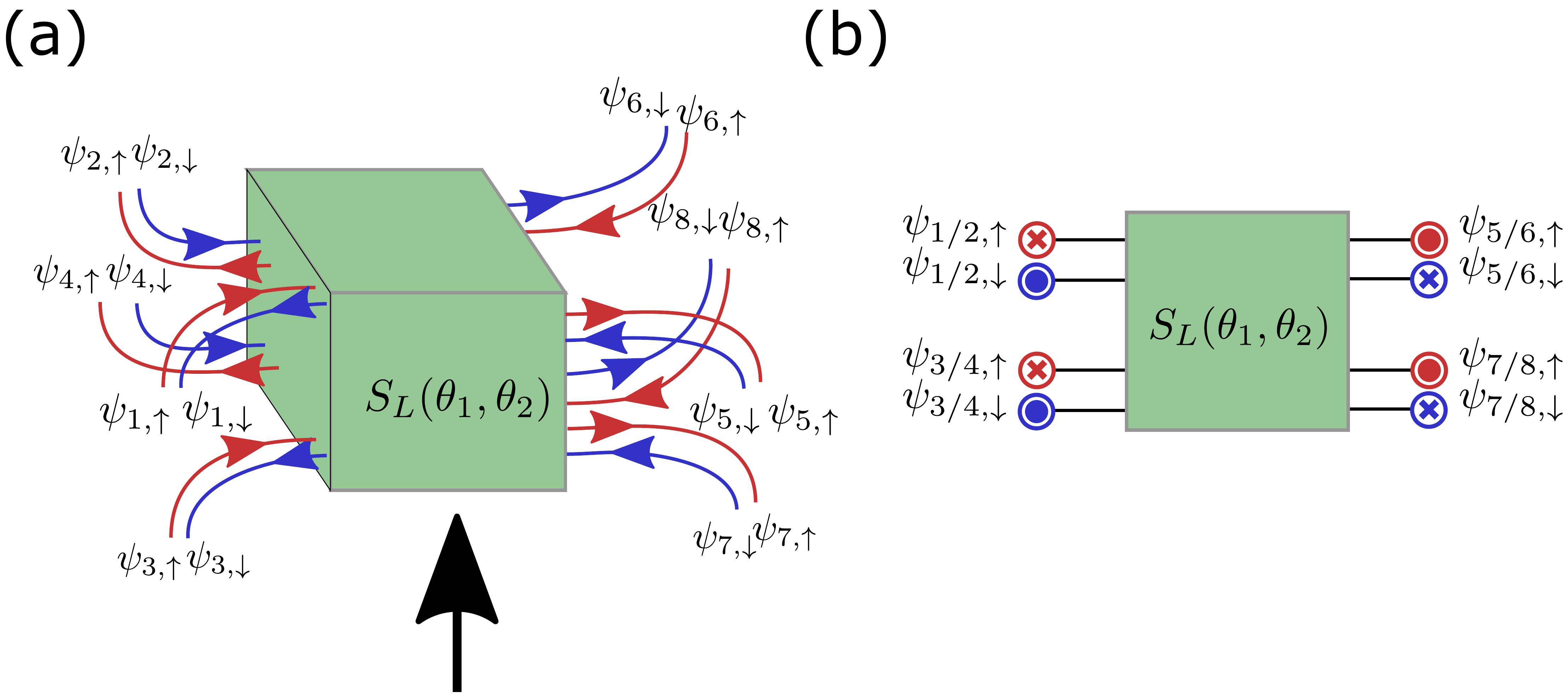}
\caption{(a) The setup and labeling of fermion modes in definition of $S_{L}(\theta_{1},\theta_{2})$ (b) The representation of $S_{L}(\theta_{1},\theta_{2})$ in the cross-sectional view (the view toward the black arrow in (a))}
\label{fig:AIIsmat3d}
\end{figure}

 The second scattering matrix $S_{L}(\theta_{1},\theta_{2})$ controls interlayer scattering. They are defined as: (See Fig.~\ref{fig:AIIsmat3d} for labeling):
\begin{equation}
\label{eq:AIIsmat3d}
\begin{split}
& \vec{\psi}_{in} = (\psi_{1, \uparrow}, \psi_{4, \downarrow}, \psi_{5, \downarrow}, \psi_{8,\uparrow}, \psi_{2, \downarrow}, \psi_{3, \uparrow}, \psi_{6, \uparrow}, \psi_{7,\downarrow}) \\ 
& \vec{\psi}_{out} = (\psi_{1, \downarrow}, \psi_{4, \uparrow}, \psi_{5, \uparrow}, \psi_{8,\downarrow}, \psi_{2, \uparrow}, \psi_{3, \downarrow}, \psi_{6, \downarrow}, \psi_{7,\uparrow}) \\
& S_{L}(\theta_{1},\theta_{2}) \vec{\psi}_{in}  = \vec{\psi}_{out}, \quad S_{L}(\theta_{1},\theta_{2}) = U_{y}^{\dagger} \begin{pmatrix}
S_{T}(\theta_{1}) & 0 \\
0 &  S_{T}(\theta_{2})
\end{pmatrix}  U_{y}	\\
&  U_{y}  =  \begin{pmatrix}
1 & 0 & 0 &0 \\
0 & 1 & 0 & 0 \\
0 & 0& 1 & 0 \\
0& 0 & 0 & 1
\end{pmatrix} \otimes \frac{1}{\sqrt{2}}\begin{pmatrix}
1 & i \\
i & 1
\end{pmatrix}
\end{split}
\end{equation}

 The above expression for $S_{L}(\theta_{1},\theta_{2})$ looks complicated, yet there is an intuitive way to  understand it. To see this, observe that $U_{y}$ mixes $\psi_{2k+1,\uparrow}$ and $\psi_{2k+2,\downarrow}$ or $\psi_{2k+1,\downarrow}$ and $\psi_{2k+2,\uparrow}$ -- it mixes two components with \textit{opposite $\sigma_{z}$ eigenvalues}. Identifying
\begin{equation}
\begin{split}
\psi_{k,\sigma_{y}=\uparrow}^{L} \sim \frac{\psi_{2k+1, \uparrow} + i \psi_{2k+2,\downarrow}}{\sqrt{2}}, \quad \psi_{k,\sigma_{y}=\downarrow}^{L} \sim \frac{\psi_{2k+1, \uparrow} - i \psi_{2k+2,\downarrow}}{\sqrt{2}}, \\
\psi_{k,\sigma_{y}=\uparrow}^{R} \sim \frac{\psi_{2k+1, \downarrow} - i \psi_{2k+2,\uparrow}}{\sqrt{2}}, \quad \psi_{k,\sigma_{y}=\downarrow}^{R} \sim \frac{\psi_{2k+1, \downarrow} + i \psi_{2k+2,\uparrow}}{\sqrt{2}},
\end{split}
\end{equation}

$L/R$ superscript account for the fact that the above equation takes linear combinations of two fermion modes that move toward left/right in Fig.~\ref{fig:AIIsmat3d}(a). In this basis that account for the basis transformation $U_{y}$, $S_{L}(\theta_{1},\theta_{2})$ is a block diagonal matrix $\begin{pmatrix} 
S_{T}(\theta_{1}) & 0 \\  
0 & S_{T}(\theta_{2}) 
\end{pmatrix}$. The idea behind the expression $S_{L}(\theta_{1},\theta_{2})$ in Eq.~\eqref{eq:AIIsmat3d} is that half of the degrees of freedom scatter according to $S_{T}(\theta_{1})$ and the remaining half according to $S_{T}(\theta_{2})$, but now the degrees of freedom are diagonal in $\sigma_{y}$ instead of $\sigma_{z}$ for other scattering matrices we have encountered so far.

\begin{figure}
\includegraphics[width=\linewidth]{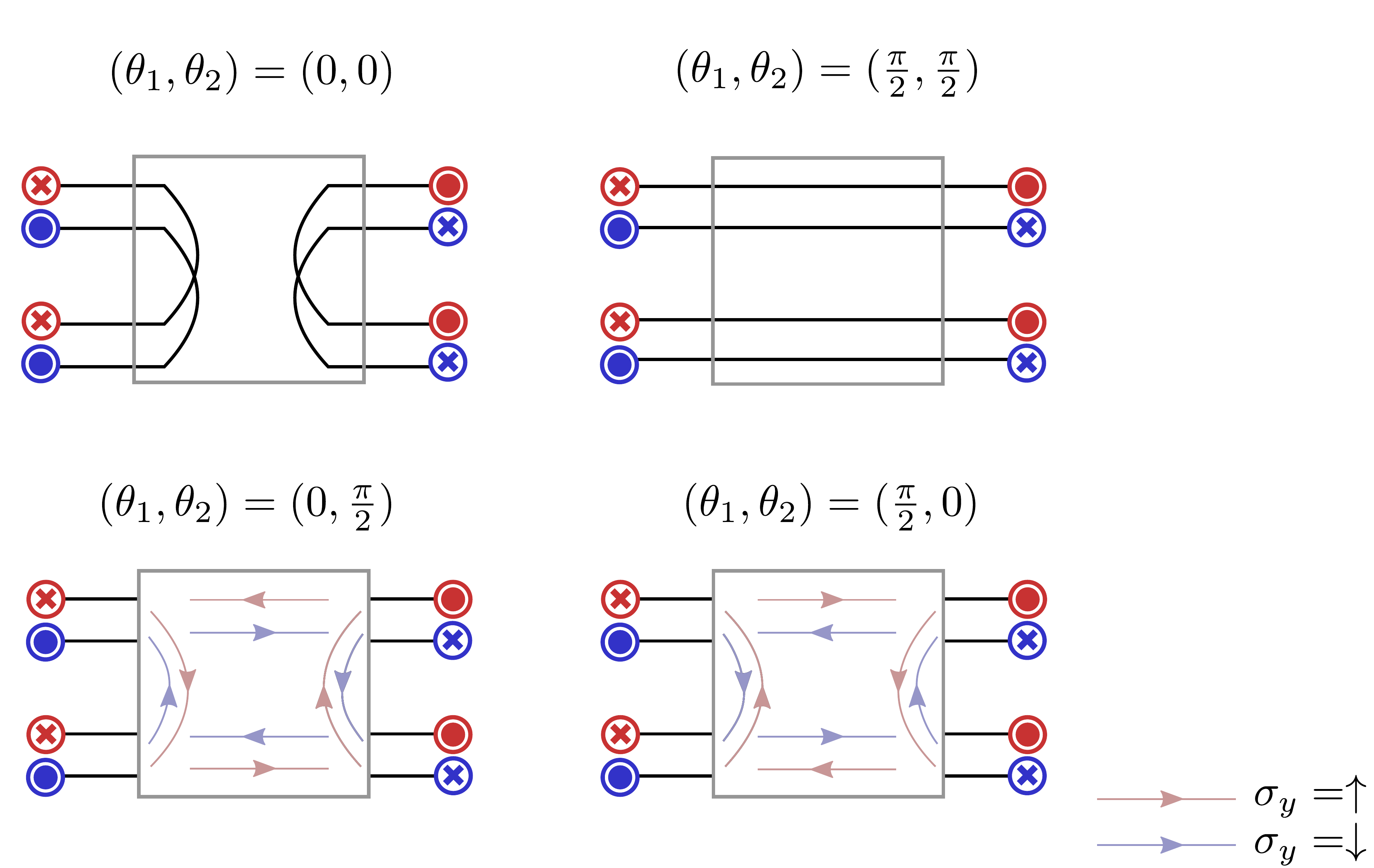}
\caption{Illustration of how the scattering matrix $S_{L}(\theta_{1},\theta_{2})$ acts in the limit $(\theta_{1},\theta_{2})=(0,0)$, $(\frac{\pi}{2}, \frac{\pi}{2})$, $(\frac{\pi}{2},0)$, and $(0,\frac{\pi}{2})$. }
\label{fig:smat3dlimit}
\end{figure}

 To give further intuition, we discuss how fermion modes scatter in special limits $(\theta_{1},\theta_{2})=(0,0)$, $(\frac{\pi}{2}, \frac{\pi}{2})$, $(\frac{\pi}{2},0)$, and $(0,\frac{\pi}{2})$. When $(\theta_{1},\theta_{2})=(0,0)$, all modes on the upper part backscatter to modes on the lower part, and vice versa (the upper left corner in Fig.~\ref{fig:smat3dlimit}). This limit maximizes interlayer backscattering. Meanwhile, in the opposite limit $(\theta_{1},\theta_{2})=(\frac{\pi}{2}, \frac{\pi}{2})$ (the upper right corner in Fig.~\ref{fig:smat3dlimit}), the modes on the upper layer and the lower layer are completely decoupled -- in this limit, there is zero interlayer scattering. $(\frac{\pi}{2},0)$, and $(0,\frac{\pi}{2})$ limit (the lower two panels in Fig.~\ref{fig:smat3dlimit}) realize more exotic scenarios in which half of the fermion modes strongly backscatter between different layers while the remaining half do not scatter to different layers at all. \textit{Note that this type of interlayer coupling is the one we would like to use to construct a strong 3D topological insulator phase, as we saw in the earlier subsection.} Whether certain fermion modes strongly scatter to the neighboring layers or stay in the same layers are dependent on $\sigma_{y}$-eigenvalues and chirality of the modes.
 
\begin{figure}
\includegraphics[width=\linewidth]{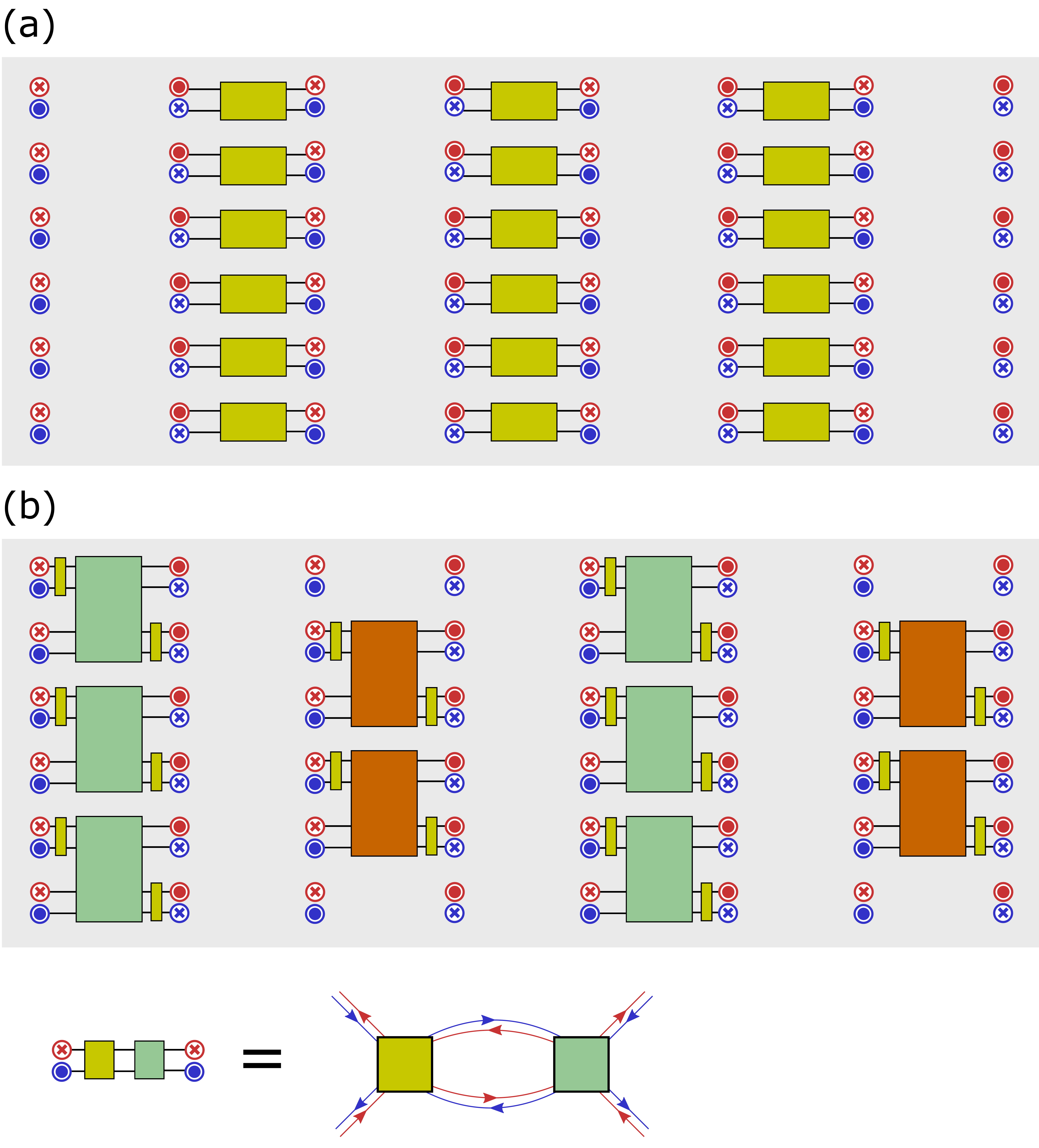}
\caption{Illustration of the cross section between (a) $z=4j+1$ and $z=4j+2$ (b) $z=4j+3$ and $z=4j+4$ showing how scattering matrices are placed in our 3D network model. $S_{T}(\frac{\pi}{4})$ are represented by gold rectangles, $S_{L}(\theta_{1},\theta_{2})$ by green rectangles, $S_{L}(\frac{\pi}{2}-\theta_{2},\frac{\pi}{2}-\theta_{1})$ by brown rectangles. In the below panel, what it means by two rectangles connected to each other with solid line: incoming/outgoing modes on the right of a gold square are identified with outgoing/incoming modes on the left of a green square.}
\label{fig:3dnetworkcross}
\end{figure}

 Now we are ready to construct the 3D network model that we will study in this paper. Our network model have two tuning parameters, $\theta_{1}$ and $\theta_{2}$. We would be primarily interested in the parameter space $0 \leq \theta_{1}, \theta_{2} \leq \frac{\pi}{2}$. There are three types of scattering matrices: $S_{L}(\theta_{1},\theta_{2})$ which we represent as green squares graphically, $S_{L}(\frac{\pi}{2}-\theta_{2},\frac{\pi}{2}-\theta_{1})$ represented by brown squares, and $S_{T}(\frac{\pi}{4})$ illustrated as gold squares. Between $z=4j+1$ and $z=4j+2$, scattering matrices are arranged as in Fig.~\ref{fig:3dnetworkcross}(a). Note that here, $S_{T}(\frac{\pi}{4})$'s simply unitarily relate modes at $(x,y)=(2i,y)$ to ones at $(x,y)=(2i+1,y)$. Scattering matrices between $z=4j+3$ and $z=4j+4$, depicted in Fig.~\ref{fig:3dnetworkcross}(b), introduce more non-trivial relations involving interlayer scattering using  $S_{L}(\theta_{1},\theta_{2})$ and $S_{L}(\frac{\pi}{2}-\theta_{2},\frac{\pi}{2}-\theta_{1})$. Finally, fermion modes $\psi_{(x,y,z=4j+2), s=\uparrow,\downarrow}$ and $\psi_{(x,y,z=4j+3), s=\uparrow,\downarrow}$ are related by random U$(1)$ phases that encode the effect of disorder, as in Eq.~\eqref{eq:AIIrandom}:
\begin{equation}
\label{eq:AIIrandom3d}
\begin{pmatrix}
e^{i\phi_{(x,y,4j+2)}} & 0 \\
0 & e^{-i\phi_{(x,y,4j+2)}}  \\
\end{pmatrix} \begin{pmatrix}
\psi_{(x,y,4j+2),\uparrow} \\
\psi_{(x,y,4j+2),\downarrow} \\
\end{pmatrix} = \begin{pmatrix}
\psi_{(x,y,4j+3),\uparrow} \\
\psi_{(x,y,4j+3),\downarrow} \\
\end{pmatrix} 
\end{equation}
Similarly, random U$(1)$ phases $e^{\pm i \phi_{(x,y,4j+4)}}$ relates fermion modes $\psi_{(x,y,4j+4),\uparrow/\downarrow}$ and $\psi_{(x,y,4j+5),\uparrow/\downarrow}$.
 
 To get some insight about the constructed network model, it is useful to consider how our network behaves in certain limits. First, let us replace both green and brown rectangles in Fig.~\ref{fig:3dnetworkcross}(b) with $S_{L}(\frac{\pi}{2},\frac{\pi}{2})$. Note that in this limit, interlayer coupling is completely absent, and the network reduces to the time-reversal symmetric network model at $\phi_{QSH}=0$, $\theta_{QSH}=\frac{\pi}{4}$ stacked along the $y$-direction. This makes it clear that our model is essentially consisted of layers, each layer hosting two 2D Dirac cones. 
 
\begin{figure}
\includegraphics[width=\linewidth]{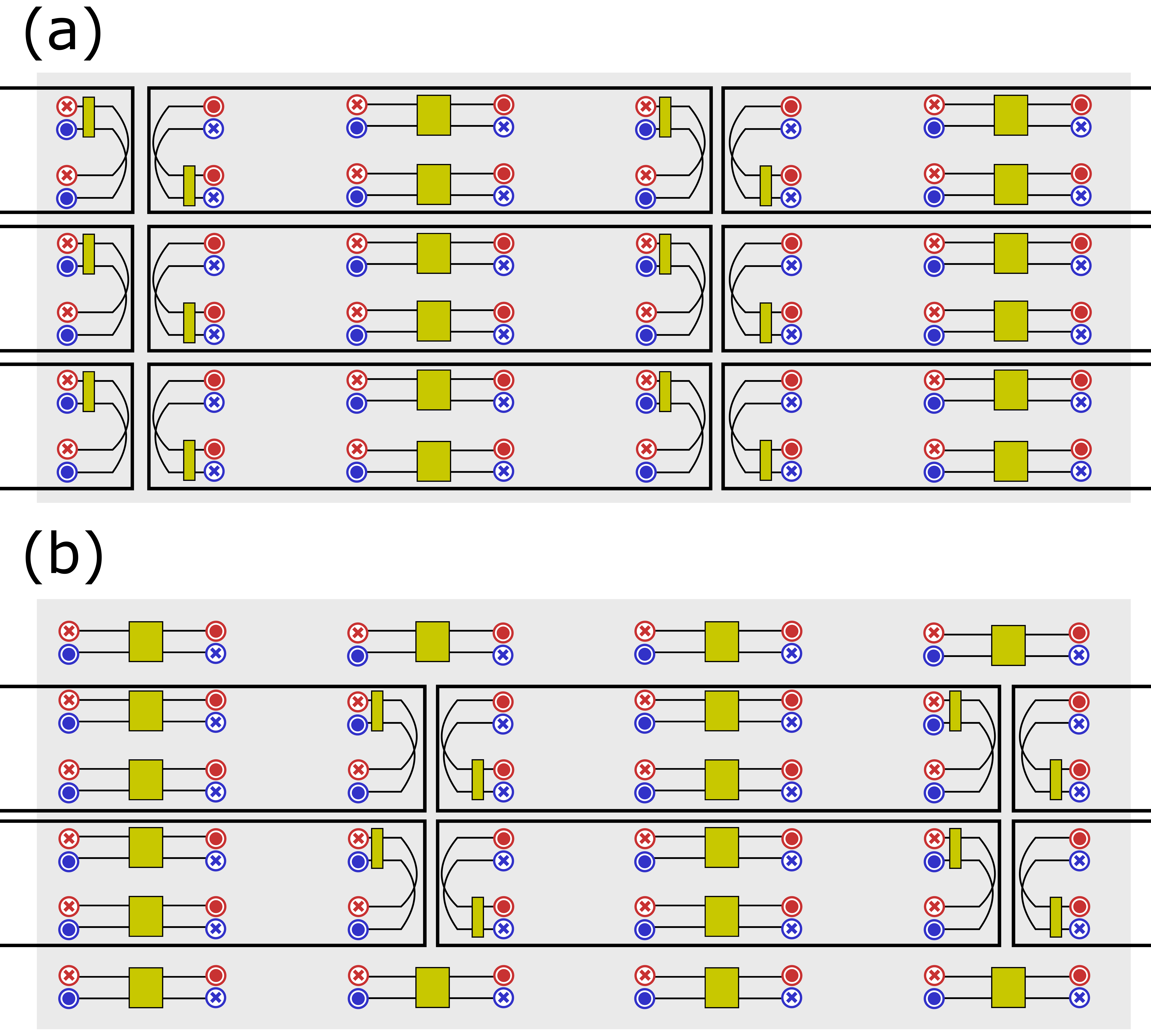}
\caption{Illustration of the cross section between $z=4j+1$ and $z=4j+2$ at (a) $(\theta_{1}, \theta_{2}) = (0,0)$ (b) $(\theta_{1}, \theta_{2}) = (\frac{\pi}{2},\frac{\pi}{2})$. In both cases, the network reduces to bundles of 1D insulating networks whose cross sections are marked as bold rectangle. In the case of $(\theta_{1}, \theta_{2}) = (\frac{\pi}{2},\frac{\pi}{2})$, when open boundary condition is imposed, the top and bottom layer decouples from the bulk and constitute surface state of a weak topological insulator.}
\label{fig:3dnetworklimit}
\end{figure}
 
 Next, we will consider $(\theta_{1}, \theta_{2}) = (0,0)$ and $(\theta_{1}, \theta_{2}) = (\frac{\pi}{2},\frac{\pi}{2})$ limits. In these limits, interlayer scattering matrices (green and brown rectangles in Fig.~\ref{fig:3dnetworkcross}(b)) reduce to either of two cases in the upper panel of Fig.~\ref{fig:smat3dlimit}
. In Fig.~\ref{fig:3dnetworklimit}(a) and (b), we pictorially represent what the cross-sectional view between $z=4j+3$ and $z=4j+4$ reduces to in these limits. The key insight is that \textit{in these limits, the network reduces to bundles of insulating 1D (in $z$-direction) networks}. Hence, near these limits, insulating phases must arise. 

Since  $(\theta_{1}, \theta_{2}) = (0,0)$ limit does not have any surface state upon imposing open boundary conditions, we identify a phase near this limit as a trivial insulator. The situation is a bit different for the limit $(\theta_{1}, \theta_{2}) = (\frac{\pi}{2},\frac{\pi}{2})$. In this case, one can explicitly see that the fermion modes at $y=1$ and $y=2L_{y}$ are decoupled from the bulk, and the fermion modes along this layer once again reduce to the time-reversal symmetric network model tuned to criticality. This surface state indicates that near $(\theta_{1}, \theta_{2}) = (\frac{\pi}{2},\frac{\pi}{2})$, our 3D network is in a \textit{weak topological insulator phase}.
 both species of fermions backscatter within their own unit cells, yielding trivial insulators. Near $(\theta_{1}, \theta_{2}) = (\frac{\pi}{2},\frac{\pi}{2})$, backscattering occurs across the neighboring unit cells. When there is a boundary normal to the $y$-direction, the layer at the boundary will be decoupled from the bulk, giving a surface state with two Dirac cones. Hence, near $(\theta_{1}, \theta_{2}) = (\frac{\pi}{2},\frac{\pi}{2})$, the system is in  \textit{a weak topological phase}. 
 
 Finally, based on the discussion in the previous subsection, near $(\theta_{1}, \theta_{2}) = (0, \frac{\pi}{2})$ and $(\frac{\pi}{2},0)$, we expect strong topological insulator phases to appear. However, these limits correspond to the cases in which interlayer scattering is, in some sense, frustrated, and we are not aware of any straightforward method to show that they are in a strong topological insulator phases. However, one can still do various quantitative analysis on the system, which we carry out in the next section, to show that 3D strong topological insulators do appear near these two limits, assuming that the network represents a sufficiently clean system.
 
 Similar to the original 2D models different parameter choices of our models can be mapped to each other by  non-local transformations. In particular, assuming periodic boundary conditions along both $x$ and $y$-directions, one can show that the following two transformations map our network model to itself:
\begin{equation}
\label{eq:dual1}
\begin{split}
& \mathcal{A}: \quad \psi_{\uparrow/\downarrow, (x,y,z)} \rightarrow \psi_{\downarrow/\uparrow, (-x,-y,z)}, \quad (\theta_{1},\theta_{2}) \rightarrow (\theta_{2},\theta_{1})\\
& \mathcal{B}: \quad \psi_{\uparrow/\downarrow, (x,y,z)} \rightarrow \psi_{\uparrow / \downarrow, (x-2,y-1,z)}, \quad (\theta_{1},\theta_{2}) \rightarrow (\frac{\pi}{2}-\theta_{2},\frac{\pi}{2}-\theta_{1})
\end{split}
\end{equation} 
 $\mathcal{A}$ is simply a spatial inversion transformation followed by a parameter change and does not change underlying topology of the network. $\mathcal{B}$ is more akin to the duality we saw in the 2D model in a sense that the non-local translation can change topological features. In particular, $\mathcal{B}$ maps a trivial insulator near $(\theta_{1},\theta_{2}) = (0,0)$ to a weak topological insulator near $(\theta_{1},\theta_{2}) = (\frac{\pi}{2}, \frac{\pi}{2})$.
 
 As a final remark on our 3D network model, our 3D model has the structure in which scattering matrices and transfer matrices along the $z$-direction can be efficiently constructed. Hence, one can use recursive methods to study properties of the 3D network model, exactly as in the 2D case.
 
\section{Phase Diagram}
\label{sec:phase}

 In the previous section, we gave intuition behind the 3D network model and explicitly constructed it. While we already know some aspects of the phase diagram of our model by considering certain limits and using the transformations in Eq.~\eqref{eq:dual1}, in this section, we will carry out a more in-depth study of the phase diagram using numerical methods. In particular, computational results in this section confirm that strong topological insulators \textit{do appear} in the parameter space as anticipated from the crude physical picture in the previous section. We will also investigate how disorder affects the phase diagram.

\subsection{System without disorder}

 Here, we are interested in phase diagram of the 	``clean network" in which we set all the random U$(1)$ phases that encapsulate the effect of disorder , i.e., $e^{i\phi_{x,y,4j+2/4j+4}}$ in Eq.~\eqref{eq:AIIrandom3d}, to $1$. Studying clean systems is much easier than studying systems with disorder for the following two reasons: First, there is no need to take any disorder averaging. Second, due to translation symmetry present in the clean network, one can study systems using a Fourier representation, which drastically reduces computational cost.
 
  All the quantities we compute in this subsection are obtained from a global scattering matrix $\mathbf{S}$ associated with a system of size $L_{x} \times L_{y} \times L_{z}$. Each unit cell has dimension $4 \times 2 \times 4$ in our model, so we define $L_{x} \times L_{y} \times L_{z}$ systems to have fermion modes $\psi_{(x,y,z),s=\uparrow/\downarrow}$ in which indices have ranges $ 1 \leq x \leq 4L_{x}$, $ 1 \leq y \leq 2L_{y}$, and $ 1 \leq z \leq 4L_{z}$. $\mathbf{S}$ relates the fermion modes at the two ends $z=1$ and $z=4L_{z}+1$. :
 \begin{equation}
\label{eq:globalS}
\mathbf{S} \begin{pmatrix}
\mathbf{\Psi}_{\text{in},z=1} \\
\mathbf{\Psi}_{\text{in},z=4L_{z}+1}
\end{pmatrix}
= \begin{pmatrix}
\mathbf{\Psi}_{\text{out},z=1} \\
\mathbf{\Psi}_{\text{out},z=4L_{z}+1}
\end{pmatrix}, \quad \mathbf{S} = \begin{pmatrix}
\mathbf{r} & \mathbf{t} \\
\tilde{\mathbf{t}} & \tilde{\mathbf{r}} \\
\end{pmatrix}
\end{equation}
 $\mathbf{r}$ and $\mathbf{t}$ refer to a reflection subblock and a transmission subblock respectively.
  
 To probe bulk physics, we will use periodic boundary condition on $x$ and $y$ direction. It is also useful to twist boundary conditions by U$(1)$ numbers $e^{ik_{x}}$ or $e^{ik_{y}}$. The twisted boundary conditions are formally defined as identifying:
\begin{equation}
\psi_{(4L_{x}+x,y,z),\uparrow/\downarrow} \equiv e^{ik_{x}}\psi_{(x,y,z),\uparrow/\downarrow}, \quad \psi_{(x,y+2L_{y},z),\uparrow/\downarrow} \equiv e^{ik_{y}}\psi_{(x,y+2L_{y},z),\uparrow/\downarrow}
\end{equation}
 We adopt notations $\mathbf{S}(k_{x},k_{y})$, $\mathbf{r}(k_{x},k_{y})$, and $\mathbf{t}(k_{x},k_{y})$ for the global scattering matrix and its subblocks obtained from twisting boundary conditions. 
 
 To investigate physics of surface states, we employ systems with open boundary conditions along the $y$ direction. This is formally implemented by replacing interlayer scattering matrices $S_{L}(\frac{\pi}{2}-\theta_{2}, \frac{\pi}{2}-\theta_{1})$  controlling backscattering between fermion modes at $y=1$ and $y=2L_{y}$ (Recall that these are brown rectangles in Fig.~\ref{fig:3dnetworkcross}(b) lying on $y=1$ and $y=2L_{y}$ ) to $S_{L}(\frac{\pi}{2}, \frac{\pi}{2})$. Since $S_{L}(\frac{\pi}{2}, \frac{\pi}{2})$ does not introduce any backscattering between two layers, the suggested replacement removes any direct coupling between $y=1$ modes and $y=2L_{y}$ modes, realizing systems with the desired boundary condition. The system is still periodic along the $x$ direction, so one may twist boundary conditions in the $x$-direction.
 
 We computed the following three quantities to map out the phase diagram:
\begin{itemize}
\item Bulk conductance across $z$ direction of the system with size $L_{x} \times L_{y} \times L_{z}$ and periodic boundary conditions in both the $x$ and $y$ directions, defined by $G_{\text{bulk}} = \text{Tr}\, \mathbf{t}^{\dagger}\mathbf{t}$. Thanks to translational symmetry,  one can compute $G_{\text{bulk}}$ equivalently from $\mathbf{S}_{u}(k_{x},k_{y})$ and $\mathbf{t}_{u}(k_{x},k_{y})$, the global scattering matrix and its transmission block for the $1 \times 1 \times L_{z}$ network with twisted boundary condition:
\begin{equation}
G_{\text{bulk}} = \sum_{n_{x}=1}^{L_{x}} \sum_{n_{y} = 1}^{L_{y}} \text{Tr} \, \left[ \mathbf{t}_{u} \left( \frac{2\pi n_{x}}{L_{x}}, \frac{2\pi n_{y}}{L_{y}} \right) \right]^{\dagger} \mathbf{t}_{u} \left( \frac{2\pi n_{x}}{L_{x}}, \frac{2\pi n_{y}}{L_{y}} \right).
\end{equation}
Computing $G_{\text{bulk}}$ from $\mathbf{t}_{u}(k_{x},k_{y})$ sidesteps computationally costly operation of directly constructing the large matrix $\mathbf{S}$ and its subblock $\mathbf{t}$.
\item $G_{\text{open}}$, conductance of the system with size $L_{x} \times L_{y} \times L_{z}$, periodic boundary condition across $x$ direction and open boundary condition across $y$ direction. $G_{\text{open}}$ probes surface states for choices of parameters in which the bulk is insulating. Similar to how we compute $G_{\text{bulk}}$, we compute $G_{\text{open}}$ from $\mathbf{S}_{ux}(k_{x})$ and $\mathbf{t}_{ux} (k_{x})$, the scattering matrix and the transmission subblock of the system which only differs from the original system by choice $L_{x}=1$ and twisted boundary condition along the $x$-direction:
\begin{equation}
G_{\text{open}} = \sum_{n_{x}=1}^{L_{x}}  \text{Tr} \, \left[ \mathbf{t}_{ux} \left( \frac{2\pi n_{x}}{L_{x}} \right) \right]^{\dagger} \mathbf{t}_{ux} \left( \frac{2\pi n_{x}}{L_{x}} \right).
\end{equation}

\item Strong topological invariant $Q$. Here, we follow the formulation from dimensional reduction previously developed in \cite{Fulga2012b}. Define $\mathbf{U}_{\mathcal{T}}$ to be an unitary part of the time-reversal symmetry action on $\mathbf{\psi_{\text{out},z=1}}$ in Eq.~\eqref{eq:globalS}. $\mathbf{U}_{\mathcal{T}}$ is analogous to $U_{\mathcal{T}}$ in Eq.~\eqref{eq:tconstraint} but generalized to act on the whole 2D arrays of modes in the cross section of our 3D network model. If $k_{x},k_{y}=0,\pi$, due to time-reversal symmetry, the reflection block satisfies,
\begin{equation}
\mathbf{r} (k_{x},k_{y}) \mathbf{U}_{\mathcal{T}} = - \mathbf{U}_{\mathcal{T}} \mathbf{r}^{T} (k_{x},k_{y}),
\end{equation}
implying that $\mathbf{r} (k_{x},k_{y}) \mathbf{U}_{\mathcal{T}}$ is skew-symmetric.
 The topological invariant associated with 3D strong topological insulators is proposed to be:
\begin{equation}
\label{eq:topoinv}
Q = Q_{k_{x} = 0} Q_{k_{x} = \pi}, \quad 
Q_{k_{x}=0,\pi} = \frac{\text{Pf} \, \mathbf{r}(k_{x},0) \mathbf{U}_{\mathcal{T}}}{\sqrt{\det \mathbf{r}(k_{x},0)}} \frac{\sqrt{\det \mathbf{r}(k_{x},\pi)}}{\text{Pf} \, \mathbf{r}(k_{x},\pi) \mathbf{U}_{\mathcal{T}}}
\end{equation}
 It is shown that $Q_{0}$ and $Q_{\pi}$ is quantized to $\pm 1$ \cite{Qi2010} in the limit that $\mathbf{r}(k_{x},k_{y})$ is a unitary matrix; this limit is valid in insulating systems with large $L_{z}$ in which transmission blocks vanish. The branch of the square root of the determinant should be chosen so that $\sqrt{\det \mathbf{r}(k_{x},k_{y})}$ should be defined continuously along the line that starts at $k_{y}=0$ and ends at $k_{y} =\pi$. Because of this, the formula requires information along the lines in the momentum space although the expression only seems to take numbers computed from time-reversal invariant momenta. 
 
 Due to translation symmetry, computing $Q$ from systems with any $L_{x}$ and $L_{y}$ contain same topological information, as long as $L_{z}$ is large enough so that the reflection block is nearly unitary. Hence, one can compute $Q$ from the system with size $1 \times 1 \times L_{z}$ in the case of the clean system.
\end{itemize}
 
\begin{figure}
\centering
\includegraphics[width = 1.0\linewidth]{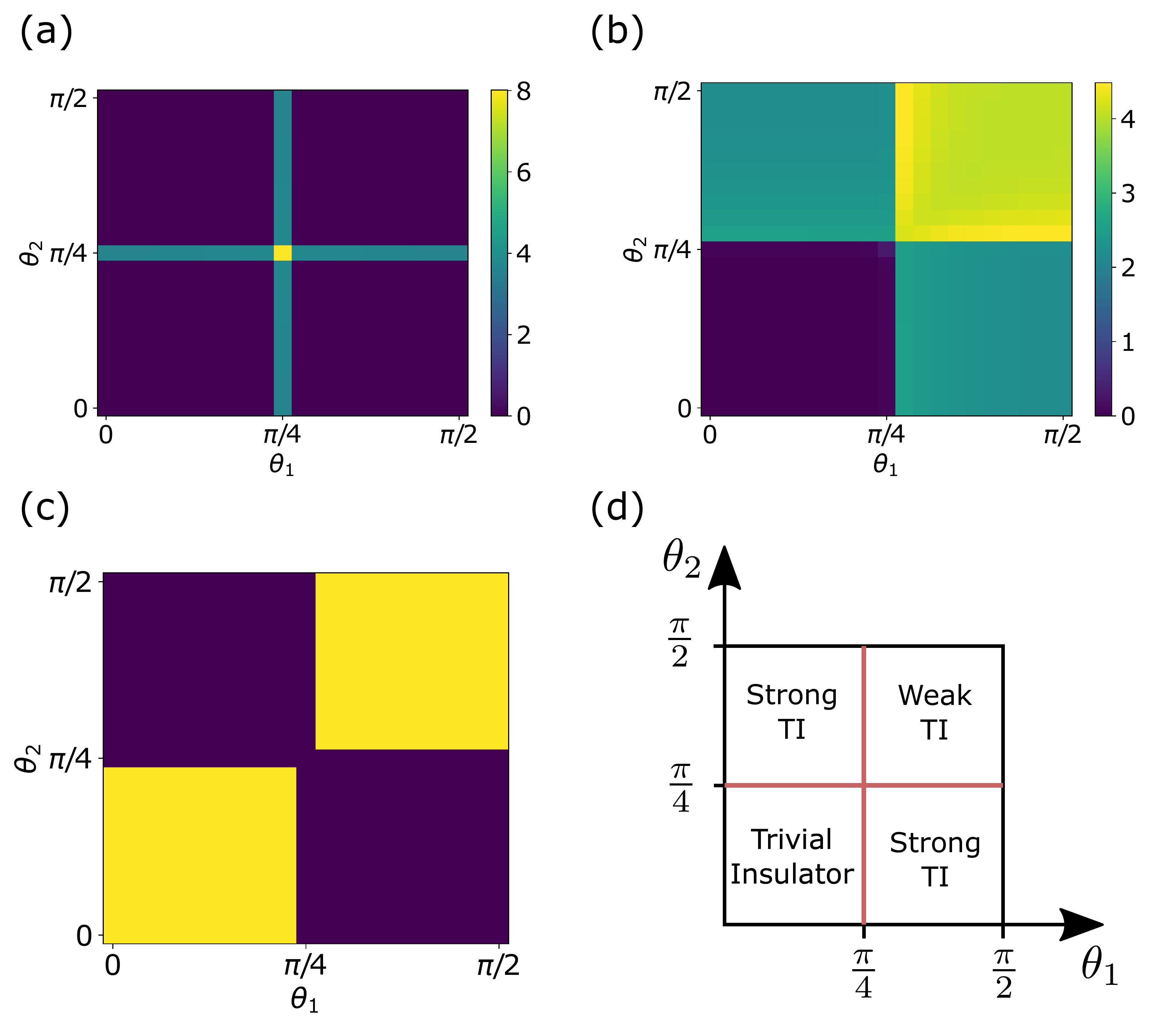}
\caption{Color plots for numerically computed values of (a) $G_{\text{bulk}}$ (b) $G_{\text{open}}$  (c) $Q$ in the two-dimensional parameter space. For the plot in (c), the yellow color stands for $Q=1$, while the purple color denotes $Q=-1$. (d) Phase diagram of the clean 3D network model based on information from the plots in (a), (b), and (c)}
\label{fig:cleanphase}
\end{figure}

 We computed these three quantities for $\theta_{1} \in \left[0, \frac{\pi}{2} \right]$ and $\theta_{2} \in \left[0, \frac{\pi}{2} \right]$, and plotted the values in Fig.~\ref{fig:cleanphase} (a,b,c). For computing $G_{\text{bulk}}$, we chose $L_{x}=L_{y}=L_{z}=256$; for computing $G_{\text{open}}$, we chose $L_{x}=L_{z}=256$, $L_{y} =32$. We fixed $L_{z}=256$ for evaluating topological invariant $Q$. 
 
 In the plot for $G_{\text{bulk}}$, $G_{\text{bulk}}$ is very close to to zero except for two lines $\theta_{1}=\frac{\pi}{4}$ and $\theta_{2} = \frac{\pi}{4}$. This indicates that the $\theta_{1}=\frac{\pi}{4}$ and $\theta_{2} = \frac{\pi}{4}$ lines correspond to the phase transition lines. $G_{\text{bulk}} \approx 4$ along the two lines, but we also point out that the number for $G_{\text{bulk}}$ at the transition lines is dependent on aspect ratios of systems in general.
 
 The two critical lines divide our parameter space $\theta_{1},\theta_{2} \in [0, \frac{\pi}{2} ]$ into four regions, each region representing a distinct insulating phase. We will refer to insulating regions around the corners $(\theta_{1},\theta_{2}) = (0,0)$, $(\frac{\pi}{2},0 )$, $(0, \frac{\pi}{2})$, and $(\frac{\pi}{2}, \frac{\pi}{2})$ as region 1,2,3, and 4 respectively. From earlier analysis, it is clear that region 1 and region 4 correspond to a trivial phase and a weak topological insulator respectively. Meanwhile, the phase separation suggests that the insulating phases in region 2 and 3 are topologically distinct from a trivial insulator in region 1 and a weak topological phase in region 4.
 
 The plots for $G_{\text{open}}$ and $Q$ make nature of the insulating phases in region 2 and 3 clearer. The insulating phase in region 1 has a vanishing $G_{\text{open}}$, signaling absence of any gapless surface state. Meanwhile, region 4 has finite conductance $G_{\text{open}} \approx 8$. We already established that this insulating phase is a weak topological insulator, so finite $G_{\text{open}}$ demonstrates the anticipated existence of the gapless surface state. Region 2 and region 3 also have finite surface conductance which is roughly \textit{half} of $G_{\text{open}}$ in the weak topological phase. This implies emergence of single Dirac cone surface states in these regions. Computing $Q$ confirms that these two insulating phases are \textit{strong topological insulators}. We drew the phase diagram based on our findings from the three quantities in Fig.~\ref{fig:cleanphase}(d). 
 
 We finish our discussion of the clean network model with implications of the transformation in Eq.~\eqref{eq:dual1} on the strong topological insulator phases in region 2 and region 3. $\mathcal{A}$ maps the network with a parameter value in region 2 to another one with a value in region 3. Since $\mathcal{A}$ involves inversion-like transformation on fermion modes, this shows that the networks in region 3 are ``mirror images" of the networks in region 2. Any statement about the strong topological insulator phase in region 2 applies identically to the phase in region 3 as well. 
 
 $\mathcal{B}$ has a more interesting consequence: $\mathcal{B}$ transforms the network in region 2 to another network in the same parameter region! Even though $\mathcal{B}$ involves highly non-local transformation, it does not transmute topology of the strong topological insulator phases. As mentioned, the same statement applies to region 3 as well. Additionally, there is a special parameter line $\theta_{1} + \theta_{2} = \frac{\pi}{2}$  \textit{invariant under the transformation} $\mathcal{B}$ in region 2 and region 3. This provides an intriguing picture of a topological phase transition in our 3D network model distinct from the 2D network models: In the 2D network models, there are dualities that connect trivial and topological insulators, and critical points/regions between two topologically distinct phases are self-dual. In our 3D network model, $\mathcal{B}$, taking a similar role to the dualities in the 2D network models, relates a trivial insulator and a \textit{weak} topological phase; strong topological insulators are intermediate phases that are mapped to themselves under $\mathcal{B}$.
  
\subsection{Adding disorder}

\subsubsection{Overview}
 
 Now, we will introduce randomness in the U(1) phases $e^{i\phi_{x,y,4j+2/4j+4}}$ in Eq.~\eqref{eq:AIIrandom3d} and see how quenched randomness modifies the phase diagram. Before embarking the numerical investigation, it is helpful to consider general consequence of adding disorder to a topological band insulator to blueprint our study. Adding small amount of disorder in general fills a band gap with electronic states. However, if disorder is sufficiently weak, these electronic states pushed to the Fermi level are localized. Hence, the system will keep its identity as a strong topological insulator although notion of the band gap disappears. In the strong disorder limit, all electronic states undergo Anderson localization, and all information about band structure is lost. Hence, we arrive at a topologically-trivial Anderson insulator.
 
 In the above consideration, the two insulators that arise in the weak disorder limit and the strong disorder limit are topologically distinct. Hence, at intermediate disorder, there should be a delocalized phase, in order to avoid a contradiction. The simplest scenario would be the existence of a critical metal on the intermediate disorder regime; this metallic phase is transformed into a strong topological insulator (a trivial insulator) upon making disorder weaker (stronger). The numerical survey on tight-binding models affirm the existence of such metallic phases. \cite{Ryu2012,Kobayashi2012}
 
 Since our interest is to investigate critical phenomena associated with strong topological insulator phases, we would like to access both strong topological insulators near the weak disorder limit and intermediate-disorder critical metal with our 3D network model. This motivates us to choose the random U$(1)$ phases $\phi$'s chosen from a normal distribution with zero mean and variance $\sigma^{2}$. This choice introduces an extra parameter $\sigma$ that tunes the disorder strength so that we can interpolate between the weak disorder limit and the intermediate disorder regime.\footnote{The conventional probability distribution of random phases in the Chalker-Coddington network model is a uniform distribution in which $\phi$'s have equal probability to take any value in $[0, 2\pi)$. In our 3D network model context, the uniform distribution corresponds to $\sigma \rightarrow \infty$ limit, and we checked that such strong disorder completely suppresses strong topological insulator phases in our 3D network model. Hence, uniform distribution is not an optimal distribution choice in our study of the 3D network model.}
  
 The above discussion also makes it evident that our primary interest in our exploration of the phase diagram is whether our network models in an insulating phase or a delocalized phase given a choice of the parameters. In a disordered system, it is not always clear whether a system is insulating or conducting just by evaluating quantities at a single system size. To determine this in a more clear and rigorous fashion, we study finite-size scaling of the localization length in a quasi-1D geometry \cite{MacKinnon1981, MacKinnon1983}. 
 
 To apply this method, consider a network with size $L \times L \times L_{z}$ under condition $L \ll L_{z}$, with periodic boundary conditions imposed on both the $x$ and $y$ directions. There is a global transfer matrix $\mathbf{T}$ that relates fermion modes at the both ends of the network. Eigenvalues of $\frac{1}{L_{z}}\ln \mathbf{T}^{\dagger} \mathbf{T}$ are associated with Lyapunov exponents along $z$-direction. There is an algorithm using QR decomposition to compute these eigenvalues efficiently and without numerical overflow \cite{MacKinnon1983}. Assuming the matrix $\frac{1}{L_{z}}\ln \mathbf{T}^{\dagger} \mathbf{T}$ does not have zero eigenvalues, its eigenvalues always come in pairs with opposite signs, $+\lambda_{i}$ and $-\lambda_{i}$. Additionally, due to the Kramers' theorem, eigenvalues always have even multiplicity. 
 
 Using the computed Lyapunov exponents, one may define a  localization length of this quasi-1D system $\xi$ as:
\begin{equation}
\xi = \frac{1}{\lambda_{\text{min}}},\quad \lambda_{\text{min}} \text{: The smallest positive eigenvalue of $\frac{1}{L_{z}}\ln \mathbf{T}^{\dagger} \mathbf{T}$}.
\end{equation}
Then, the idea is to utilize a dimensionless parameter $\Lambda = \frac{\xi}{L}$ that takes a dimension of the cross section $L$ into account and its finite-size scaling beta function $\beta =\frac{d \ln \Lambda}{d \ln L}$. The sign of the beta function determines whether the system is insulating or conducting in the thermodynamic limit. In an insulating phase, $\beta < 0$, while in a metallic phase, $\beta > 0$. $\beta =0$ implies scale invariance and is associated with a critical point. In this subsection, as a quick diagnosis of signs of the beta functions, we compute and compare $\Lambda$'s for $L=6,8,10$. All quantities in this subsection are obtained from systems with length $L_{z} = L \times 10^{4}$, averaged over four disorder realizations.
 
 One may also probe the topology of the disordered network in the insulating phases by computing the strong topological invariant Q defined in Eq.~\eqref{eq:topoinv} and studying (de-)localization properties of the network model with open boundary conditions. However, we will see that all localized phases we encounter in the disordered network model are adiabatically connected to some insulating phases in the clean limit. Thus, it is unnecessary to compute these numbers in the dirty systems directly.

\subsubsection{Numerical Result}

 In our numerical exploration of the network model with U$(1)$ phases generated from a normal distribution, we fix $\theta_{2} =0$ and tune $\theta_{1}$ to be in vicinity of $\frac{\pi}{4}$. For convenience, we will occasionally use $\delta_{1} = \theta_{1} - \frac{\pi}{4}$ instead of $\theta_{1}$ for notational simplicity.
  
\begin{figure}
\includegraphics[width=1.0\linewidth]{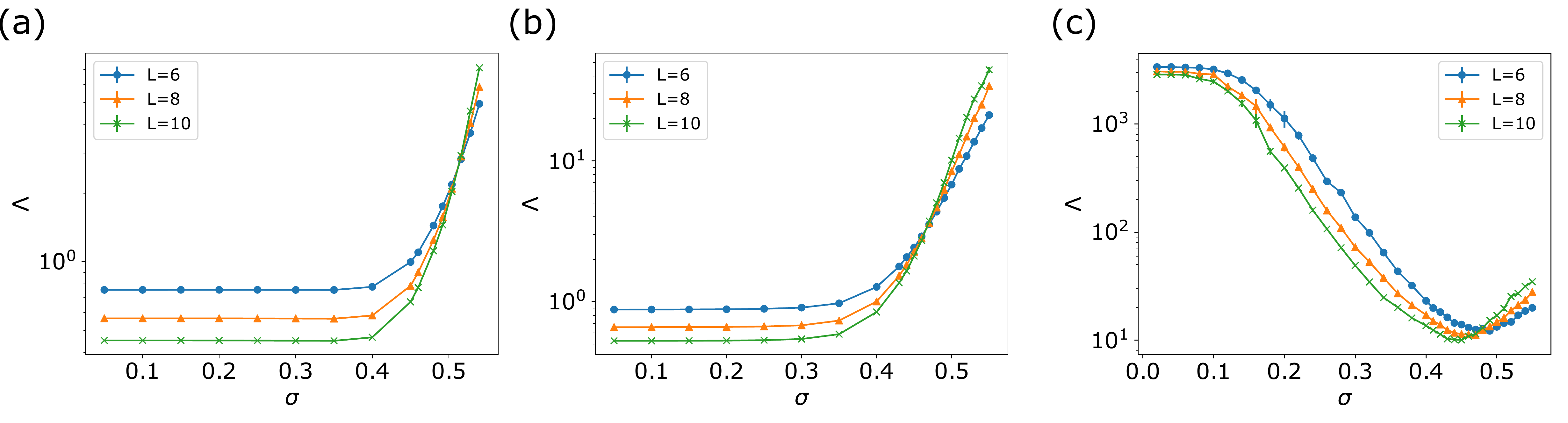}
\caption{Plots of numerically computed $\Lambda$'s versus different values of disorder strength $\sigma$, upon fixing $\theta_{2}=0$ and  setting (a) $\delta_{1} = 0.30$ (b) $\delta_{1} = -0.30$ (c) $\delta_{1} = 0.0$}
\label{fig:renorvscan}
\end{figure}
 
 First, to observe the disorder-tuned metal-insulator transition, we choose $ \delta_{1} $  from $\{-0.3,-0.2,$ $-0.1,0.0,0.1,0.2,0.3 \}$. For each choice of $\delta_{1}$, we sweep through different values of $\sigma$ to inspect how the finite-size scaling properties of $\Lambda$ change. For all $\delta_{1}$ choices except for $\delta_{1} = 0$, we confirmed that upon increasing $\sigma$, the network model undergoes insulator-metal transitions. Plots of $\Lambda$ for $\delta_{1} = \pm 0.3$ in Fig.~\ref{fig:renorvscan} (a) and (b) showcase this behavior, manifesting crossover from $\beta <0$  in the small disorder limit to $\beta > 0$ in larger disorder.
 
 In all $\delta_{1}$'s we investigated except for $\delta_{1} = 0$, the insulating phases in the weak disorder regime are adiabatically connected to the insulators in the network with the same $(\theta_{1},\theta_{2})$ and zero disorder. Therefore, as promised earlier, topology of the insulating, dirty 3D network model is naturally inherited from its clean counterpart. Especially, the observed insulator-metal transition at $\delta_{1} > 0$ corresponds to the topological phase transition from a strong topological insulator phase to a metal. 
 
 The plot for $\delta_{1} = 0$ in Fig.~\ref{fig:renorvscan} (c) is more puzzling. In the clean limit, this value of the parameter corresponds to a critical point between the strong topological insulator and the trivial insulator. However, the plot suggests that $\beta < 0$ in a wide range of small but non-zero disorder, implying that upon adding quenched randomness the network turns into an insulator at $\delta_{1} = 0$. Upon increasing $\sigma$, the metallic phase with $\beta >0$ appears here as well. 
 
  Another peculiar feature at $\delta_{1} = 0$ is that while there are clear regions where $\beta >0$ and $\beta <0$, it is difficult to pinpoint a critical point between the two regions where $\beta$ is supposed to be zero. At the putative critical point, $\beta=0$ implies $L$-independence of $\Lambda$, but the curves for $L=6$, $L=8$, and $L=10$ do not meet nicely at one point. Such ``mismatch" commonly appears in numerical analysis of $\Lambda$ in disordered systems \cite{MacKinnon1994,Wang1996,Slevin1999} and is attributed to irrelevant variable effect. This effect is also expected to be present in data for Fig.~\ref{fig:renorvscan}(a) and (b) upon increasing precision of data but particularly afflicts data plotted in Fig.~\ref{fig:renorvscan}(c) severely. We comment on this large irrelevant variable effect at $\delta_{1}=0$ near the end of this section.
  
\begin{figure}
\includegraphics[width=1.0\linewidth]{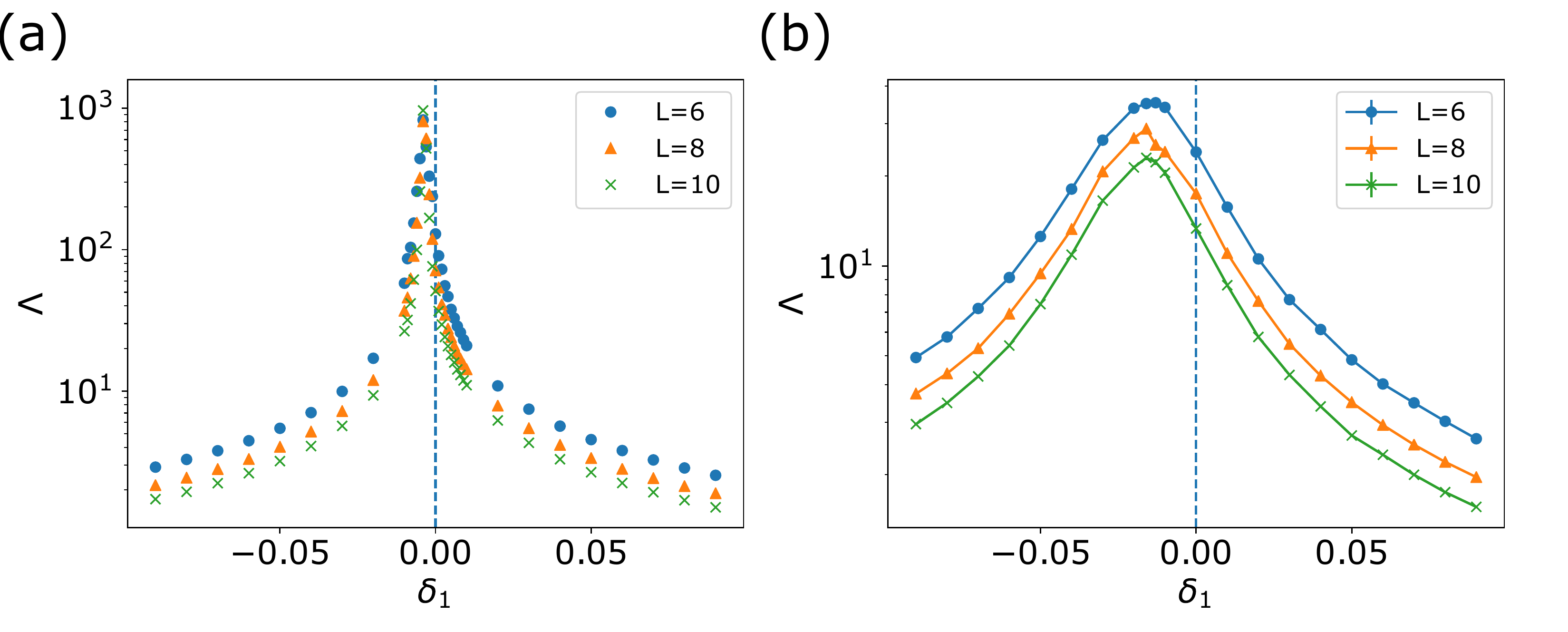}
\caption{Plots of numerically computed $\Lambda$ versus different values of $\delta_{1}$, upon fixing $\theta_{2}=0$ and setting (a) $\sigma= 0.30$ (b) $\sigma = 0.40$}
\label{fig:renorhscan}
\end{figure}

  To reveal the nature of the insulating phase at $\delta_{1} = 0$ with finite disorder, we fix $\sigma = 0.30, 0.35,$ $0.40$ and scan through different values of $\delta_{1} \in [-0.1,0.1]$. The plot at $\sigma =0.30$ in Fig.~\ref{fig:renorhscan}(a) shows that there is a point in which $\Lambda$ is roughly independent of $L$ surrounded by $\beta<0$ regimes. The regime with $\beta <0$ on the left is adiabatically connected to a trivial insulator and the $\beta < 0$ regime on the right to a strong topological insulator. Interestingly, the $\beta = 0$ critical point is \textit{shifted away} from $\delta_{1} = 0$. Hence, we see insulating behaviors at $\delta_{1} = 0$ with finite disorder simply because the transition point between the strong topological insulator phase and the trivial insulator is renormalized away from $\delta_{1} = 0$. Since the insulator at $\delta_{1} =0$ is adiabatically connected to the insulator that appears at $\delta_{1}>0$, it is identified as a strong topological insulator. We see a similar behavior for $\sigma = 0.35$ as well. \footnote{Here, we implicitly assumed that the phase transition from a trivial insulator to a topological insulator is direct when disorder is weak. This assumption has been believed to be true for the following reason: Typically, a 3D Dirac cone emerges at a critical point between a strong topological insulator and a trivial insulator in band theory context. Since 3D Dirac cone is perturbatively stable against disorder, one may argue that the aforementioned direct transition is generically possible. However, it was recently proposed that non-perturbative rare-region effect actually destabilizes Dirac semimetal \cite{Nandkishore2014,Pixley2016b}. This implies that any strength of disorder turns the Dirac semimetal critical point into a diffusive metallic region, forbidding any direct transition between a strong topological insulator and a trivial insulator. This issue is not focus of our work, and this rare region effect is likely to be inaccessible from the small system numerics in this paper. Hence, we will presume that there is a direct transition between the strong topological phase insulator and the trivial insulator phase in our network model at weak disorder throughout this paper. However, we do not rule out possibility that the ``critical point" governing the phase transition between the two insulating phases actually belongs to a metallic region. }

 The plot for $\sigma =0.40$ in  Fig.~\ref{fig:renorhscan}(b) is perplexing -- it seems that $\beta < 0$ everywhere, without any region/point with $\beta \geq 0$. However, from our numerics of scanning through different $\sigma$'s at fixed $\delta_{1}$, the two insulating phases at $\delta_{1} = \pm 0.1$ have different topology. Hence, the absence of any region with $\beta \geq 0$ poses an apparent contradiction.

 We do not have a clear explanation at this point why it seems to be $\beta < 0$ everywhere at $\sigma = 0.4$. We have one guess: This unexpected behavior may come from the fact that this point is in the vicinity of a tricritical point \cite{Goswami2011,Kobayashi2014,Syzranov2015,Roy2014,Pixley2016a} in which the three phase transition lines -- the line between the strong topological insulator and the metallic phase, the line between the trivial insulator and the metallic phase, and the line governing the direct transition between the strong topological insulator and the trivial insulator  -- meet. The one-parameter finite-size scaling ansatz of the renormalized length $\Lambda$ is probably not valid near this point. We believe that the large irrelevant field effect in Fig.~\ref{fig:renorvscan}(c) can be also traced back to the same reason.
 \begin{figure}
\includegraphics[width=0.5\linewidth]{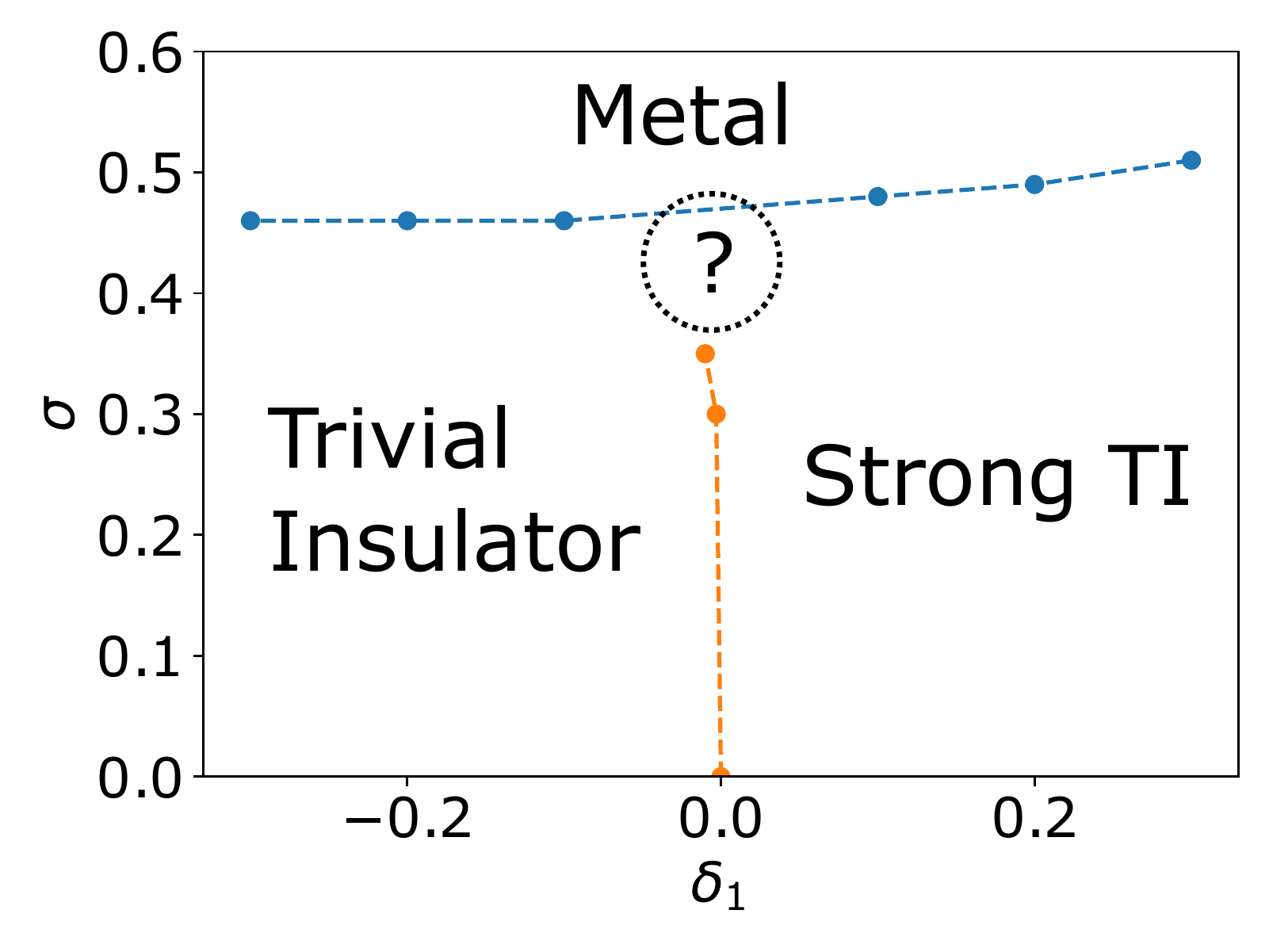}
\caption{Estimated phase boundaries of our 3D network model upon fixing $\theta_{2} =0$ and varying $(\delta_{1} = \theta_{1} - \frac{\pi}{4}, \sigma)$.}
\label{fig:phaseD}
\end{figure}
 
 We present the phase diagram obtained from our numerical data in Fig.~\ref{fig:phaseD}. Blue dots are estimated from fixing $\delta_{1} = \theta_{1} - \frac{\pi}{4}$ to certain values and sweeping through $\delta$ and finding the value of $\sigma$ with minimum of $\left| \ln \left( \frac{\Lambda_{L=10}}{\Lambda_{L=8}}\right) \right|$, taken as a proxy for $|\beta|$. Orange dots are obtained similarly by fixing $\sigma$ and scanning through $\delta_{1}$. Near the area where the metal-insulator boundary and the direct strong TI-trivial insulator boundary should meet, as we stated earlier, estimating the phase boundaries is difficult, and their location is left as blank in a question-marked region in Fig.~\ref{fig:phaseD}.

\section{Critical Exponent from Finite-size Scaling Analysis}

 Here, we estimate a critical exponent for the strong topological insulator-metal transition using the finite-size scaling properties of the renormalized length $\Lambda$. Here, we follow finite-size scaling framework developed in \cite{MacKinnon1994,Slevin1999} incorporating corrections from irrelevant variables. In this section, we fix $\delta_{1} = 0.3$ and $\theta_{2} =0$ and study strong topological insulator to metal transition tuned by $\sigma$ in this work.
 
 If we follow one-parameter scaling of $\Lambda$ near a metal-insulator transition reviewed in the earlier section, at the critical point $\sigma_{c}$, $\Lambda$ is independent of the system size. One can also formulate a simple ansatz of how $\Lambda$ behaves near critical point with a critical exponent $\nu$ associated with scaling of the correlation length $\xi \sim \frac{1}{|\sigma-\sigma_{c}|^{\nu}}$.  $\Lambda$, or equivalently $\log \Lambda$, is a dimensionless number and hence we assume it takes a one-parameter scaling form,
\begin{equation}
\label{eq:scale1}
\log \Lambda(\sigma,L) = f\left( (\sigma-\sigma_{c})L^{1/\nu} \right)
\end{equation}
 In addition, it is natural to expect that $\Lambda(\sigma,L)$ is an analytic function of $\sigma$ for any finite $L$. Hence, one may Taylor-expand $\log \Lambda(\sigma,L)$ near $\sigma_{c}$ as:
\begin{equation}
\log \Lambda(\sigma,L) = \log \Lambda_{c} + A_{01} (\sigma-\sigma_{c})L^{1/\nu}  +  A_{02} (\sigma-\sigma_{c})^{2}L^{2/\nu} + \cdots
\end{equation}
where $\Lambda_{c}$ is defined as $\Lambda$ at the criticality.

 In numerical practices, $\sigma$ is a UV variable which includes potentially infinite number of irrelevant variables in IR. While the  effect of irrelevant variables should vanish in the thermodynamic limit, in numerical calculation of $\Lambda$, due to the small system sizes, this effect will inadvertently show up in data, most commonly in a manner that points with $\frac{\partial \Lambda (\sigma, L)}{\partial L} =0$ seem to shift systematically to one direction as $L$ increases; recall that this effect appeared in our earlier studies of the phase diagram of our 3D network model. To accommodate this effect, we introduce $y$, the smallest scaling dimension of irrelevant scaling variables, and some function of $\sigma$, $g(\sigma)$, which represent the dominant irrelevant scaling variable. Then, one can modify the ansatz in Eq.~\eqref{eq:scale1} as:
\begin{equation}
\log \Lambda(\sigma,L) = f\left( (\sigma-\sigma_{c})L^{1/\nu}, \frac{g(\sigma)}{L^{y}} \right)
\end{equation}
Starting from the above modified ansatz, once again one may assume analyticity and Taylor-expand, $g$ first then $f$. In this section, we assume $g(\sigma)$ is a constant and will fit our numerical data to the following formula:
\begin{equation}
\label{eq:ansatzfinal}
\begin{split}
\log \Lambda(\sigma,L) & \approx \log \Lambda_{c} + A_{01} (\sigma-\sigma_{c})L^{1/\nu}  +  A_{02} (\sigma-\sigma_{c})^{2}L^{2/\nu} \\ 
& \quad \quad + L^{-y} \left[ A_{10} + A_{11} (\sigma-\sigma_{c})L^{1/\nu}  +  A_{12} (\sigma-\sigma_{c})^{2}L^{2/\nu} \right] 
\end{split}
\end{equation} 

\begin{figure}
\includegraphics[width=0.5\linewidth]{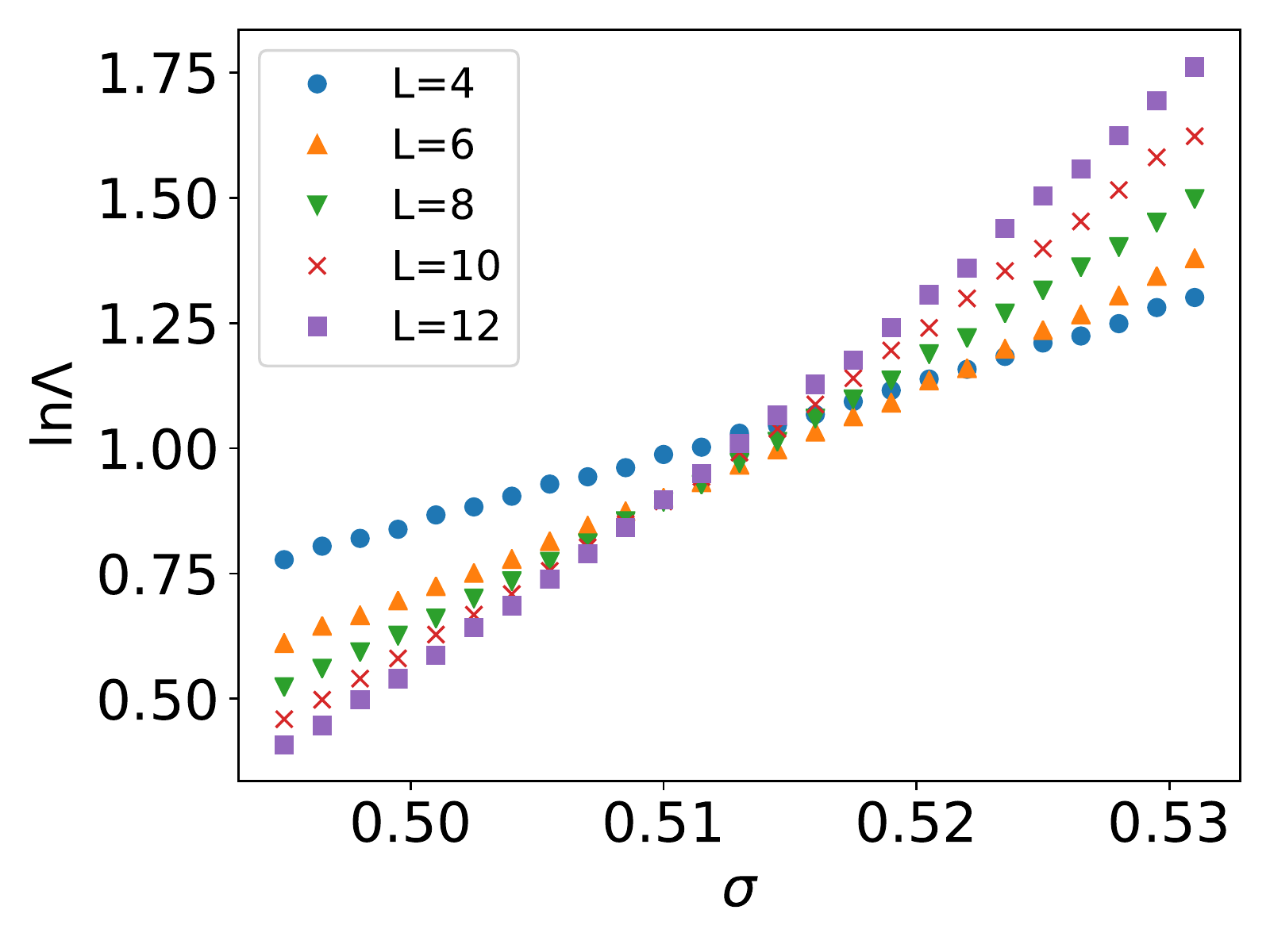}
\caption{Plots of numerically computed $\Lambda$ we used for extracting the critical exponent $\nu$.}
\label{fig:renor_FS}
\end{figure}

 For the non-linear fitting, we numerically computed $\Lambda$ from the network with size $L \times L \times L_{z}$ for $L=4,6,8,10,12$, $L_{z}= 10000 \times L$ for $L\neq 12$ and $L_{z} = 8000 \times L$ for $L=12$, taking disorder averaging for $20$ independent realization. The estimated uncertainty in data points range from $0.17 \% - 0.69\%$. We plotted numerical estimations of $\Lambda$'s we will use for the fitting in Fig.~\ref{fig:renor_FS}. There are 25 data points for each $L$, total of 125 data points. The plot clearly shows the irrelevant variable effect in $L=4$ and $L=6$ data.
 
  We could obtain a fit with reasonable stability, checked by varying initial condition of the non-linear fit and Monte-Carlo resampling. The fitting result yields: (95$\%$ confidence interval on error estimate):
\begin{equation}
\begin{split}
\nu = 1.311 \pm 0.033 \\
 y = 3.244 \pm 0.538 \\
 \sigma_{c} = 0.508 \pm 0.001 \\ 
\log \Lambda_{c} = 0.838 \pm 0.023 \\
\chi^{2} = 109.3
\end{split}
\end{equation}
 We emphasize that finite-size scaling analysis was almost impossible in tight-binding models \cite{Ryu2012} because a small band gap on the strong TI side, or alternatively large correlation length on the insulating side, corrupts behavior of the renormalized length $\Lambda$ near the critical point in small system sizes severely. Interestingly, this problem is absent in the network model.
 
 Another important point to note from our fitting result is that the estimated $\nu$ is \textit{not compatible} with symplectic Anderson transition exponent computed from (non-topological) tight-binding models in \cite{Asada2005}, at least in 95$\%$ confidence interval. This is also at odds with the conjecture in \cite{Ryu2012} that the strong topological insulator-metal transition belongs to the same universality class to the plain-vanilla symplectic Anderson transition. While it is tempting to claim a new universality class based on our result, we believe that our data and fitting are rather crude. Hence, we advise readers to take results in this section rather as demonstration that it is feasible to extract critical exponents from the finite-size scaling analysis using our network model than as a quantitative prediction. More detailed numerical studies involving data with smaller error bars and evaluations of critical exponents along different lines in the parameter space are needed to make any definite conclusion about the universality classes.
 
\section{Conclusion}
 
 In this paper, we constructed the 3D network model that exhibits both weak and strong time-reversal symmetric topological insulator phases. Our model, similar to the original 2D network model by Chalker and Coddington, is amenable to numerical methods using quasi-1D geometries; this enables extensive numerical survey of our model. Through numerics, we observed that some insulating phases, especially the strong topological phases, in the 3D network model transform into metallic phases upon adding sufficiently strong quenched randomness. This result is consistent with previous results based on tight-binding Hamiltonians. 
 
  Our model also has novel features that were not previously appreciated from the analysis based on tight binding models: First, we found that in our network model, there is a non-local transformation that relates a weak topological insulator phase and a trivial phase. Strong topological insulators are intermediate phases between the two aforementioned phases that are mapped to itself under the transformation. Second, we showed that through finite-size scaling analysis one can extract the critical exponent of the strong topological insulator-metal transition. In the previous approach using tight-binding models technical difficulties associated with band gap scales complicated such analysis.
  
  There are several future directions. The most immediate extension of our work will be improving our estimates on the critical exponent for the strong topological insulator-metal transition by enhancing precision and computing exponents for different choices of parameters. This will allow one to make definite conclusions about whether the topological insulator-metal transition is in the same universality class as the sympletic Anderson transition without any topological feature. 
  
  It would be also interesting to extend our construction to other topological phase transitions in three spatial dimensions. For instance, the field-theory description of time-reversal invariant topological superconductor is largely similar to that of time-reversal invariant topological insulators, except the complex fermions in the latter are replaced by real Majorana fermions. Hence, we expect that there is a fairly simple generalization of our model to the 3D topological superconductor transition. In contrast, a 3D network model for AIII topological insulator transitions poses a more interesting challenge since there is no intrinsic 2D AIII topological insulator and hence no weak 3D AIII topological insulator. Our approach inadvertently includes weak topological insulator phases in the phase diagram. This leads us to believe that designing the AIII 3D network model involves more non-trivial modification of our methodology. 
  
  Finally, we found the non-local transformation under which the strong topological insulator phases are invariant, but we did not comment on physical origin of this transformation. Connecting this non-local transformation to physical dualities/symmetries would be helpful in understanding topological phase transitions in greater depth.

\section*{Acknowledgement}
We thank Jason Alicea, Jing-Yuan Chen, Andrew Essin, Prashant Kumar, Gil Refael, and Xiao-Qi Sun for discussion. This work was
supported in part by the US Department of Energy,
Office of Basic Energy Sciences, Division of Materials
Sciences and Engineering, under contract number DEAC02-76SF00515

\begin{appendices}
\section{More on Scattering Matrices and Transfer Matrices}
\begin{figure}
\centering
\includegraphics[width=0.7\linewidth]{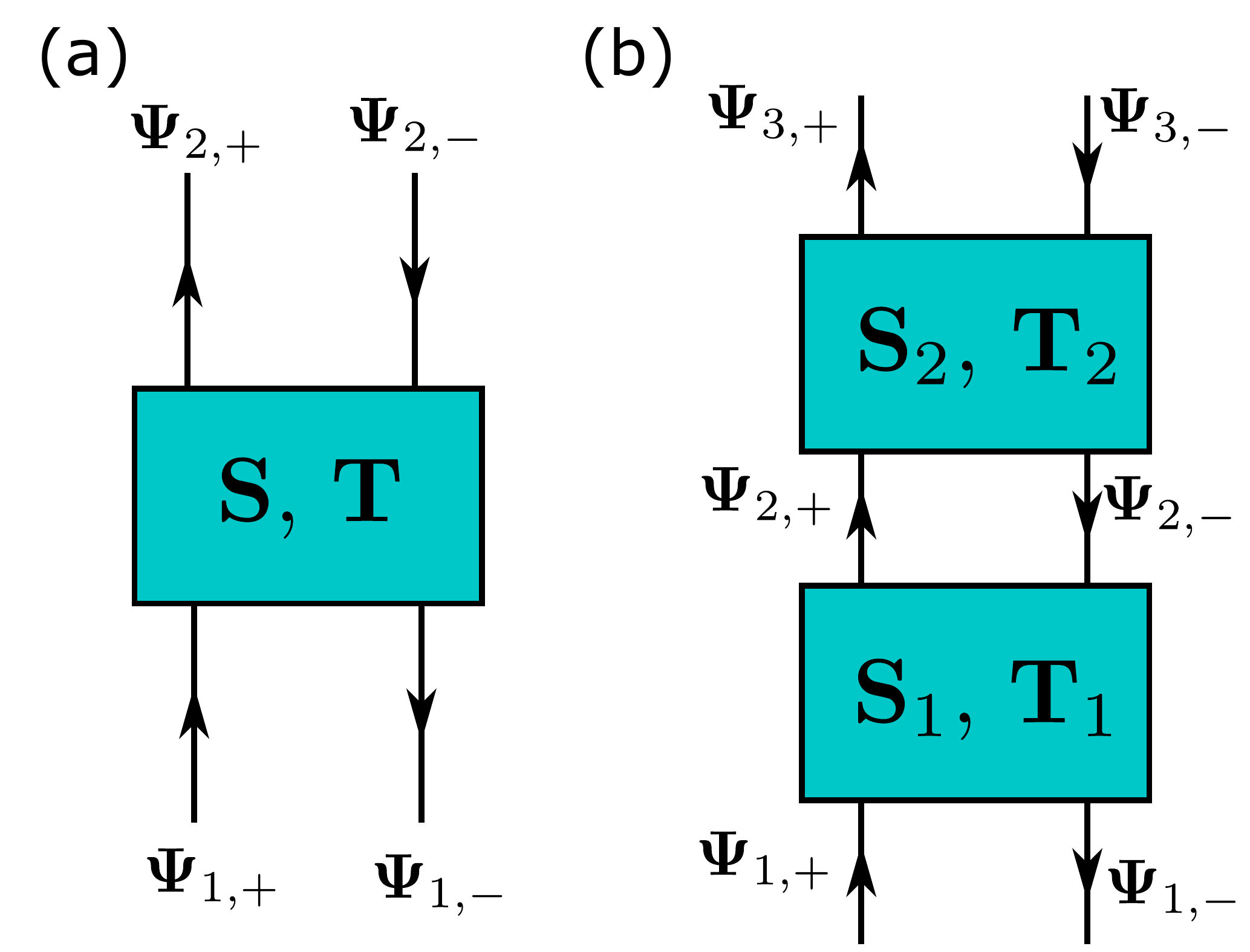}
\caption{Graphical depiction of notations used in (a)  formulas involving a single scattering/transfer matrix and (b) formulas involving two scattering/transfer matrices .}
\label{fig:AppAsetup}
\end{figure}

 In this appendix, we present formulas involving scattering matrices and transfer matrices used for the numerical implementation. We also discuss how these formulas can be used for constructing a global scattering matrix in practice.
 
\subsection{Conversion formula between scattering matrices and transfer matrices}
 
 In the main text, when reviewing the 2D network model and constructing our 3D network model, we formulated the models using scattering matrices. Meanwhile, in numerical practices, one commonly uses transfer matrices. Here, we give formula on how scattering matrices and transfer matrices are mathematically related. Generally a single scattering matrix and a transfer matrix take the following form:
\begin{equation}
\label{eq:stsingledef}
\begin{split}
& \mathbf{S} \begin{pmatrix}
\mathbf{\Psi}_{1,+} \\ 
\mathbf{\Psi}_{2,-}
\end{pmatrix} = \begin{pmatrix}
\mathbf{\Psi}_{1,-} \\ 
\mathbf{\Psi}_{2,+}
\end{pmatrix}, \quad \mathbf{S} = \begin{pmatrix}
A & B \\
C & D
\end{pmatrix} \\
& \mathbf{T} \begin{pmatrix}
\mathbf{\Psi}_{1,+} \\ 
\mathbf{\Psi}_{1,-}
\end{pmatrix} = \begin{pmatrix}
\mathbf{\Psi}_{2,+} \\ 
\mathbf{\Psi}_{2,-}
\end{pmatrix}, \quad \mathbf{T} = \begin{pmatrix}
a & b \\
c & d
\end{pmatrix}
\end{split}
\end{equation}
$\mathbf{\Psi}_{1,+}$, $\mathbf{\Psi}_{1,-}$, $\mathbf{\Psi}_{2,+}$, and $\mathbf{\Psi}_{2,-}$ are length-$n$ vectors, and each block of $\mathbf{S}$ and $\mathbf{T}$ matrix is a $n \times n $ matrix. This setup is illustrated in Fig.~\ref{fig:AppAsetup}(a). Each block of a scattering matrix can be expressed in terms of transfer matrix subblocks, and vice versa, as the following:
\begin{equation}
\label{eq:strel}
\mathbf{S} = \begin{pmatrix}
-d^{-1} c & d^{-1}\\
a - bd^{-1}c & b d^{-1} 
\end{pmatrix}, \quad \mathbf{T} = \begin{pmatrix}
C - DB^{-1} A & D B^{-1} \\
-B^{-1}A & B^{-1}\\
\end{pmatrix}
\end{equation}
 Above two equations allow one to convert a scattering matrix to a transfer matrix and vice versa.

\subsection{Formula involving two scattering matrices}
 
 Now, we will consider a setup illustrated in Fig.~\ref{fig:AppAsetup}(b) in which there are two scattering nodes described by scattering matrices $\mathbf{S}_{1}$ and $\mathbf{S}_{2}$, or equivalently two transfer matrices $\mathbf{T}_{1}$ and $\mathbf{T}_{2}$. We label fermion modes and matrix subblocks analogously to Eq.~\eqref{eq:stsingledef}, as the following: 
\begin{equation}
\begin{split}
& \mathbf{S}_{1} \begin{pmatrix}
\mathbf{\Psi}_{1,+} \\ 
\mathbf{\Psi}_{2,-}
\end{pmatrix} = \begin{pmatrix}
\mathbf{\Psi}_{1,-} \\ 
\mathbf{\Psi}_{2,+}
\end{pmatrix}, \quad \mathbf{S}_{1} = \begin{pmatrix}
A_{1} & B_{1} \\
C_{1} & D_{1}
\end{pmatrix} \\
& \mathbf{S}_{2} \begin{pmatrix}
\mathbf{\Psi}_{2,+} \\ 
\mathbf{\Psi}_{3,-}
\end{pmatrix} = \begin{pmatrix}
\mathbf{\Psi}_{2,-} \\ 
\mathbf{\Psi}_{3,+}
\end{pmatrix}, \quad \mathbf{S}_{2} = \begin{pmatrix}
A_{2} & B_{2} \\
C_{2} & D_{2}
\end{pmatrix} 
\end{split}
\end{equation}
 Our goal is to construct a formula for a scattering matrix/transfer matrix of the combined system $\mathbf{S}_{c}$/$\mathbf{T}_{c}$ that gives relation between $\mathbf{\Psi}_{1,\pm}$ and $\mathbf{\Psi}_{3,\pm}$:
\begin{equation}
\mathbf{S}_{c} \begin{pmatrix}
\mathbf{\Psi}_{1,+} \\ 
\mathbf{\Psi}_{3,-}
\end{pmatrix} = \begin{pmatrix}
\mathbf{\Psi}_{1,-} \\ 
\mathbf{\Psi}_{3,+}
\end{pmatrix}, \quad \mathbf{T}_{c} \begin{pmatrix}
\mathbf{\Psi}_{1,+} \\ 
\mathbf{\Psi}_{1,-}
\end{pmatrix} = \begin{pmatrix}
\mathbf{\Psi}_{3,+} \\ 
\mathbf{\Psi}_{3,-}
\end{pmatrix}
\end{equation}
 Building $\mathbf{T}_{c}$ is easy -- due to multiplicative structure of transfer matrices, $\mathbf{T}_{c}$ is simply equal to $\mathbf{T}_{2}\mathbf{T}_{1}$. Meanwhile, $\mathbf{S}_{c}$ can be built from subblocks of $\mathbf{S}_{1}$ and $\mathbf{S}_{2}$ according to the following formula:
\begin{equation}
\label{eq:sc1}
\mathbf{S}_{c} = \begin{pmatrix}
A_{1} + B_{1} (I - A_{2} D_{1})^{-1}A_{2}C_{1} & B_{1}(I - A_{2} D_{1})^{-1} B_{2} \\
C_{2}C_{1} + C_{2}D_{1}(I - A_{2} D_{1})^{-1} A_{2}C_{1} & D_{2} + C_{2}D_{1} (I - A_{2} D_{1})^{-1} B_{2}
\end{pmatrix}
\end{equation}
 Note that above formula requires one to take matrix inverse, an operation more costly than matrix multiplication. Hence, constructing $\mathbf{T}_{c}$ with matrix multiplication is generally faster than building $\mathbf{S}_{c}$ from the above formula. As an alternative method, if one knows $\mathbf{T}_{2}^{-1}$(note the inverse) and $\mathbf{S}_{1}$, $\mathbf{S}_{c}$ may be computed from subblocks of $\mathbf{T}_{2}^{-1}$ and $\mathbf{S}_{1}$ according to the following formula:
\begin{equation}
\label{eq:sc2}
\begin{split}
&\mathbf{T}_{2}^{-1} = \begin{pmatrix}
\overline{a}_{2} & \overline{b}_{2} \\ 
\overline{c}_{2} & \overline{d}_{2} 
\end{pmatrix}, \quad 
 \mathbf{S}_{c} = \begin{pmatrix}
A_{c} & B_{c} \\
C_{c} & D_{c}
\end{pmatrix} \\
& C_{c} = (\overline{a}_{2} - D_{1}\overline{c}_{2})^{-1} C_{1} \\
& D_{c} = (\overline{a}_{2} - D_{1}\overline{c}_{2})^{-1} (D_{1}\overline{d}_{2} - \overline{b}_{2}) \\ 
& A_{c} = A_{1} + B_{1}\overline{c}_{2} C_{c}\\
& B_{c} =  B_{1}\overline{c}_{2} D_{c} + B_{1}\overline{d}_{2}
\end{split}
\end{equation}

 The above expression also requires matrix inversion. However, it has a smaller number of matrix multiplications than Eq.~\eqref{eq:sc1} and may be evaluated more efficiently.

\subsection{Constructing a global scattering matrix in practice}

Based on the result from the previous two subsections, we now consider a setting in which we have fermion modes $\mathbf{\Psi}_{i,\pm}$, $i = 1,2, \cdots n+1$ living on $i$th slice of the system and scattering matrices $\mathbf{S}_{j}$ ($j=1,2,\cdots n$) that endow unitary relations between fermion modes $\mathbf{\Psi}_{j,\pm}$ and $\mathbf{\Psi}_{j+1,\pm}$. We will show how to construct a global scattering matrix $\mathbf{S}$ that relates $\mathbf{\Psi}_{1,\pm}$ and $\mathbf{\Psi}_{n+1,\pm}$, the fermion modes that reside on the both ends of the system in numerical practice. We assume that cost of converting $\mathbf{S}_{j}$ to $\mathbf{T}_{j}$ or $\mathbf{T}_{j}^{-1}$ is negligible; this is valid in the network model context since scattering matrices across two neighboring slices are often composed of smaller blocks due to locality. Hence, one may convert scattering matrices and transfer matrices to each other by applying the formula Eq.~\eqref{eq:strel} to the individual blocks. Numerical cost of such operations is suppressed compared to matrix inversion or matrix multiplication operations applied to the full-sized transfer/scattering matrices when system size is large. 

 If one considers computational cost only, the most efficient method is to compute a global transfer matrix $\mathbf{T} = \mathbf{T}_{n} \mathbf{T}_{n-1}\cdots\mathbf{T}_{1}$ and convert $\mathbf{T}$ to $\mathbf{S}$. However, this method does not work in practice, due to the following reason: $\mathbf{T}$, especially when the system length $n$ is large, often has very large eigenvalues that lead to numerical overflow. Alternatively, one may start from $\mathbf{S}_{1}$ and iteratively apply Eq.~\eqref{eq:sc1} to $\mathbf{S}_{2},\mathbf{S}_{3}, \cdots,\mathbf{S}_{n}$ or Eq.~\eqref{eq:sc2} to $\mathbf{T}_{2}^{-1},\mathbf{T}_{3}^{-1}, \cdots,\mathbf{T}_{n}^{-1}$. While this method is free from overflow, this method is clearly slower than multiplying transfer matrices.
 
 In numerical implementation, it is most practical to use the hybrid method -- use matrix multiplications to construct (inverse) transfer matrices across several slices, number of slices small enough to avoid overflow, and apply Eq.~\eqref{eq:sc2} iteratively to construct a global scattering matrix $\mathbf{S}$ from the transfer matrices for multi-slice chunks.
\end{appendices}

\bibliography{3dnetwork}{}
\bibliographystyle{utphys}
\end{document}